\documentclass[12pt]{article}

\usepackage{geometry}
\usepackage[utf8]{inputenc}
\usepackage{indentfirst}
\usepackage{graphicx}
\usepackage{algorithm}
\usepackage{algpseudocode}
\usepackage{amsmath, amsfonts}   
\usepackage{longtable}
\usepackage{booktabs}
\usepackage{multirow}
\usepackage{threeparttable}
\usepackage[figuresright]{rotating}
\usepackage{tikz}
\usepackage{doi}
\usepackage{url}
\usepackage{hyperref}
\usetikzlibrary{arrows,graphs,chains,decorations.pathreplacing,positioning,shapes.geometric,automata,arrows.meta}
\usepackage{subfig}
\usepackage{smartdiagram}
\usepackage{natbib}
\usepackage[figuresright]{rotating}
\bibpunct[, ]{(}{)}{,}{a}{}{,}%

\title{\bfseries The Dial-a-Ride Problem with Synchronized Visits}

\author{
	Boshuai Zhao\\
	{\small Université Paris-Saclay, CentraleSupélec, Laboratoire Génie Industriel, France}\\
	{\small \texttt{boshuai.zhao@centralesupelec.fr; zhaoboshuai1995@163.com}}
	\and
	Jakob Puchinger\\
	{\small EM Normandie Business School, Métis Lab, France}\\
	{\small Université Paris-Saclay, CentraleSupélec, Laboratoire Génie Industriel}\\
	{\small \texttt{jakob.puchinger@centralesupelec.fr}}
	\and
	Roel Leus\\
	{\small ORSTAT, Faculty of Economics and Business, KU Leuven, Belgium}\\
	{\small \texttt{roel.leus@kuleuven.be}}
}

\begin{document}
	\maketitle
	
	
		
		
	
	\textbf{Abstract:} The limited capacity of drones and future one- or two-seat modular vehicles requires multiple units to serve a single large customer (i.e., a customer whose demand exceeds a single vehicle's capacity) simultaneously, whereas small customers (i.e., those whose demand can be served by a single vehicle) can be consolidated in one trip. This motivates the Dial-a-Ride Problem with Synchronized Visits, where a fleet of drones must be routed and scheduled to transport orders at minimum cost. We propose four formulations: arc-based, event-based, time-space event-based (TSEF), and time-space fragment-based (TSFrag). An event is defined as a tuple of a location and a set of onboard customers, while a fragment represents a partial path. For TSEF and TSFrag, we also employ the dynamic discretization discovery (DDD) algorithm, which iteratively refines an initial low-resolution time-space network to obtain a continuous-time optimal solution. Computational results show that the event-based formulation performs best under low request intensity (few customers per unit time), whereas TSFrag with DDD excels with high request intensity; both substantially outperform the arc-based formulation. When implemented with DDD, TSFrag also requires less time and fewer iterations than TSEF\@. 
        We also apply our methods to the classical dial-a-ride problem, where we find that that TSFrag with DDD can replace callbacks in case of high request intensity, and that using DDD
        is more beneficial to this problem than to the pickup-and-delivery problem with time windows.


	
	\section{Background}\label{ch3:background}
	
\begin{figure}[H]
	\centering
	\begin{minipage}[t]{0.33\textwidth}
		\centering
		\includegraphics[width=\textwidth]{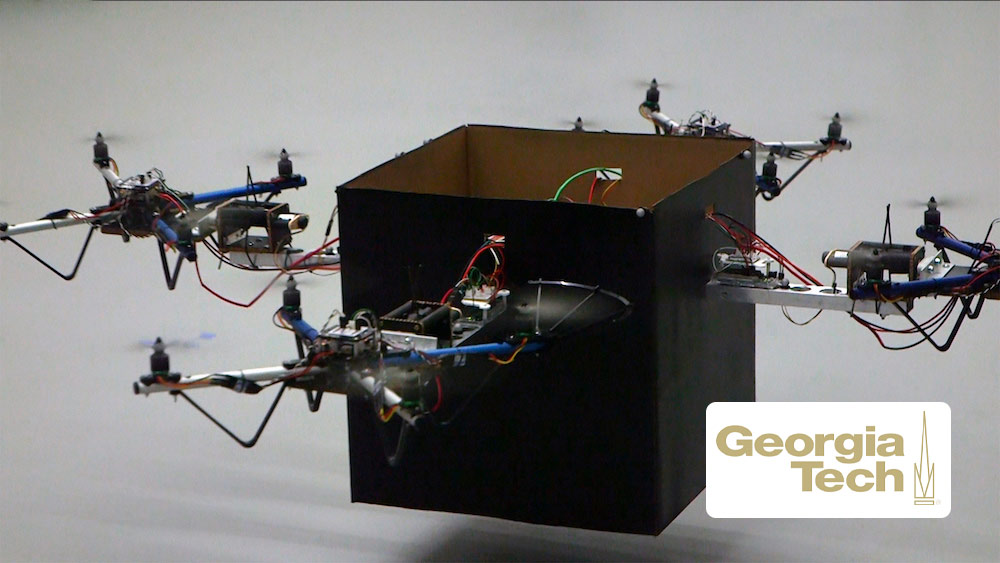}
		\label{fig:multidrones}
	\end{minipage}\hfill
	\begin{minipage}[t]{0.33\textwidth}
		\centering
		\includegraphics[width=\textwidth]{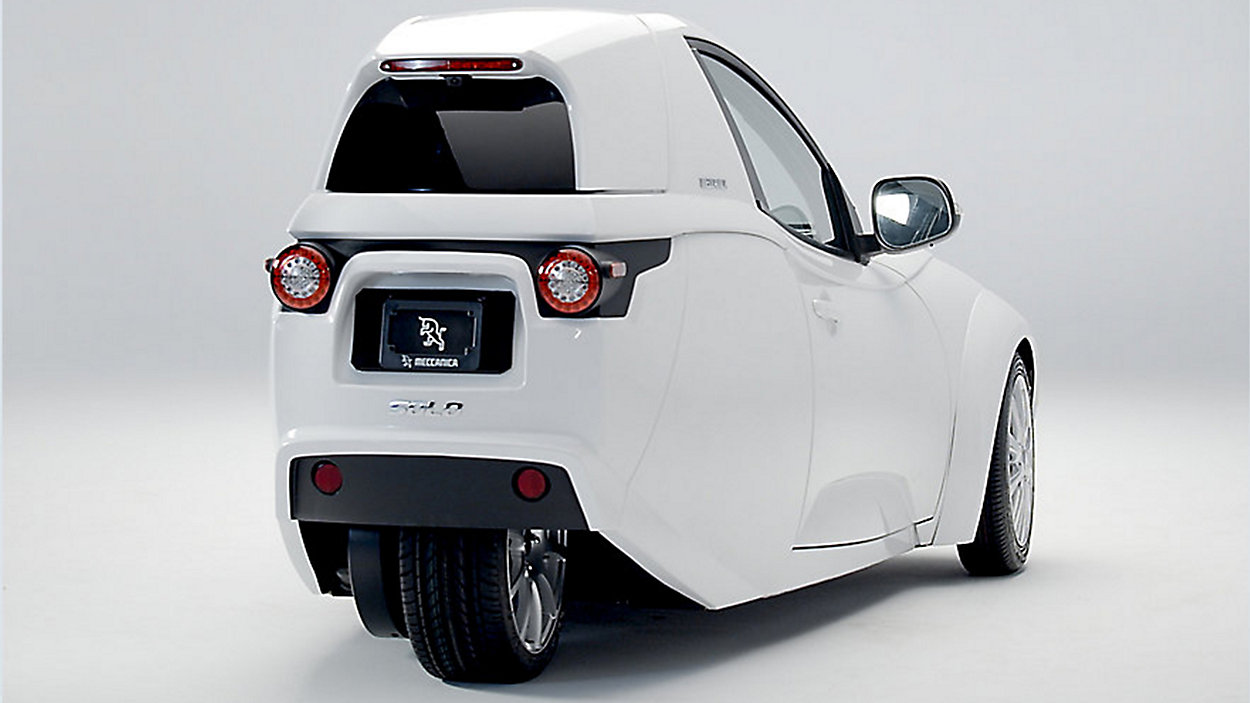}
		\label{fig:singlevehicle}
	\end{minipage}\hfill
	\begin{minipage}[t]{0.292\textwidth}
		\centering
		\includegraphics[width=\textwidth]{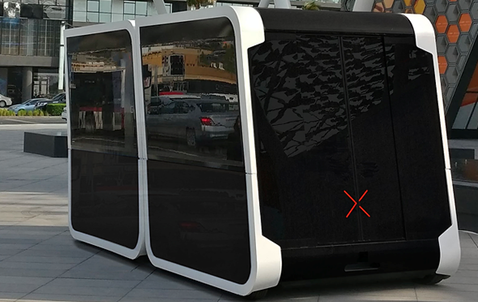}
		\label{fig:modularmulti}
	\end{minipage}
	\caption{Illustrations of multi-drone delivery, a single-seat vehicle, and a modular vehicle}
	\label{fig:allfig}
\end{figure}

	Drone delivery is increasingly adopted in pickup-and-delivery applications such as food distribution and emergency response \citep{Dukkanci2024}. A common setting is the dial-a-ride problem (DARP), where a fleet of vehicles (drones) is routed to pick up and deliver customers at minimum routing cost. Recent extensions allow multiple drones to cooperate. While a single drone can serve several \textit{small customers} (each with demand within a single drone's delivery capacity) concurrently \citep{MENG2024685}, it cannot handle \textit{large customers} whose demand alone exceeds this capacity. In such cases, multiple drones must synchronize their operations to jointly serve a large customer \citep{Nguyen23}.  When these drones are mechanically coupled to a common payload, as in Fig.~\ref{fig:allfig}, synchronized departure and arrival become a hard operational requirement.
	
	A similar situation arises in passenger transport, where a large customer may represent multiple passengers exceeding a single vehicle's capacity. Although currently uncommon, the rise of single-seat and modular autonomous vehicles (see Fig.~\ref{fig:allfig}) is expected to shift demand toward smaller configurations \citep{Monica}, making synchronized operations increasingly relevant. In a study for 15 American cities, \citet{FHWA2021} found that  64.2\% of private ride-hailing trips involved a single passenger, 21.5\% of the trips were made by two passengers, and 13.0\% by three or more passengers. These figures imply that one- or two-seat autonomous vehicles could serve the majority of mobility needs. Such low-capacity designs would offer significant energy advantages over conventional four-seat vehicles originally intended for private household use. Consequently, low-capacity vehicles may prevail, and synchronized operations among such vehicles could accommodate the occasional demand for group or family travel in future shared-mobility systems.

\begin{figure}[H]
	\centering
	
	\begin{minipage}{0.45\textwidth}
		\centering
		\includegraphics[scale=0.3]{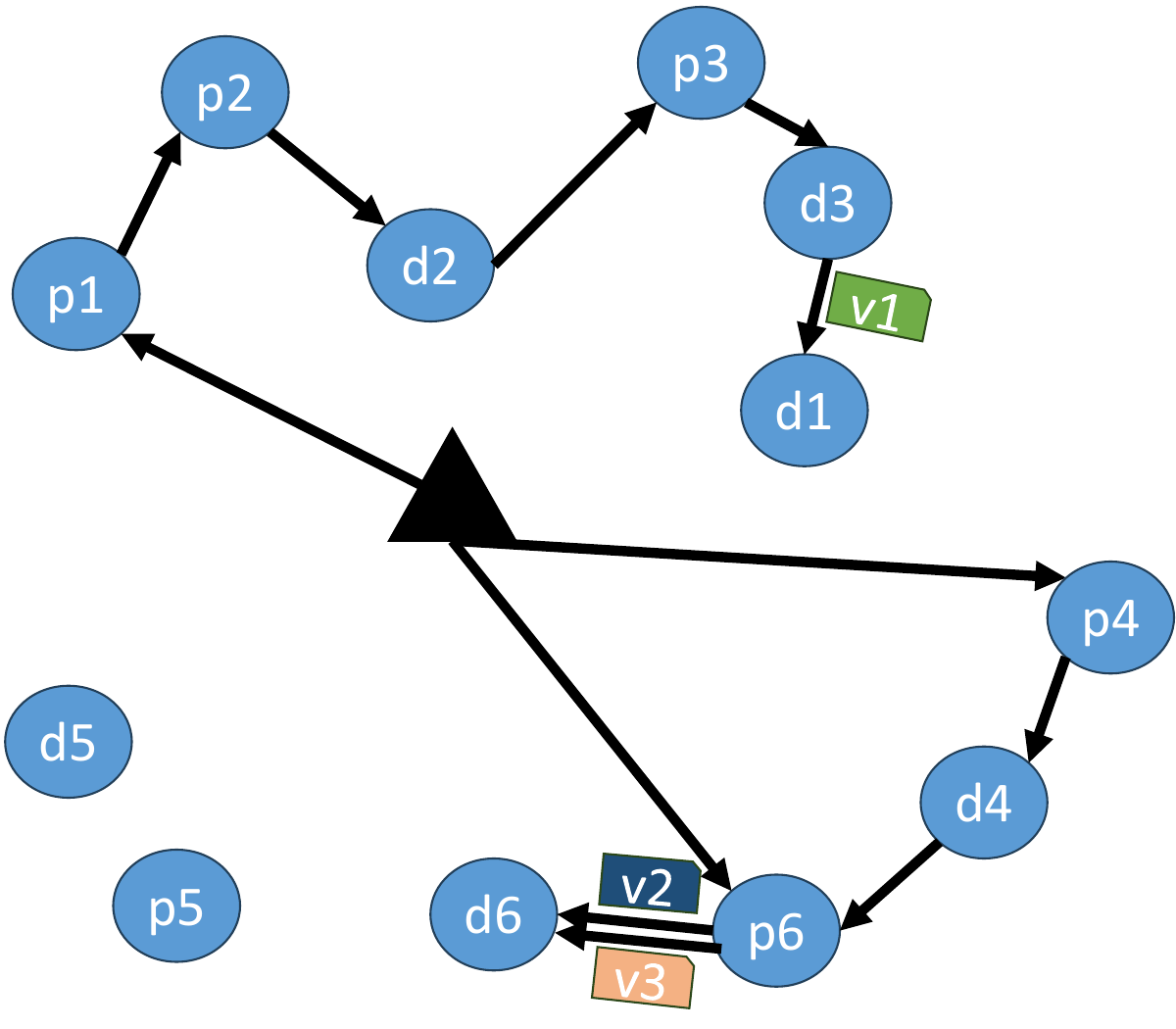}
	\end{minipage}\hspace{15pt}
	\begin{minipage}{0.45\textwidth}
		\centering
		\includegraphics[scale=0.3]{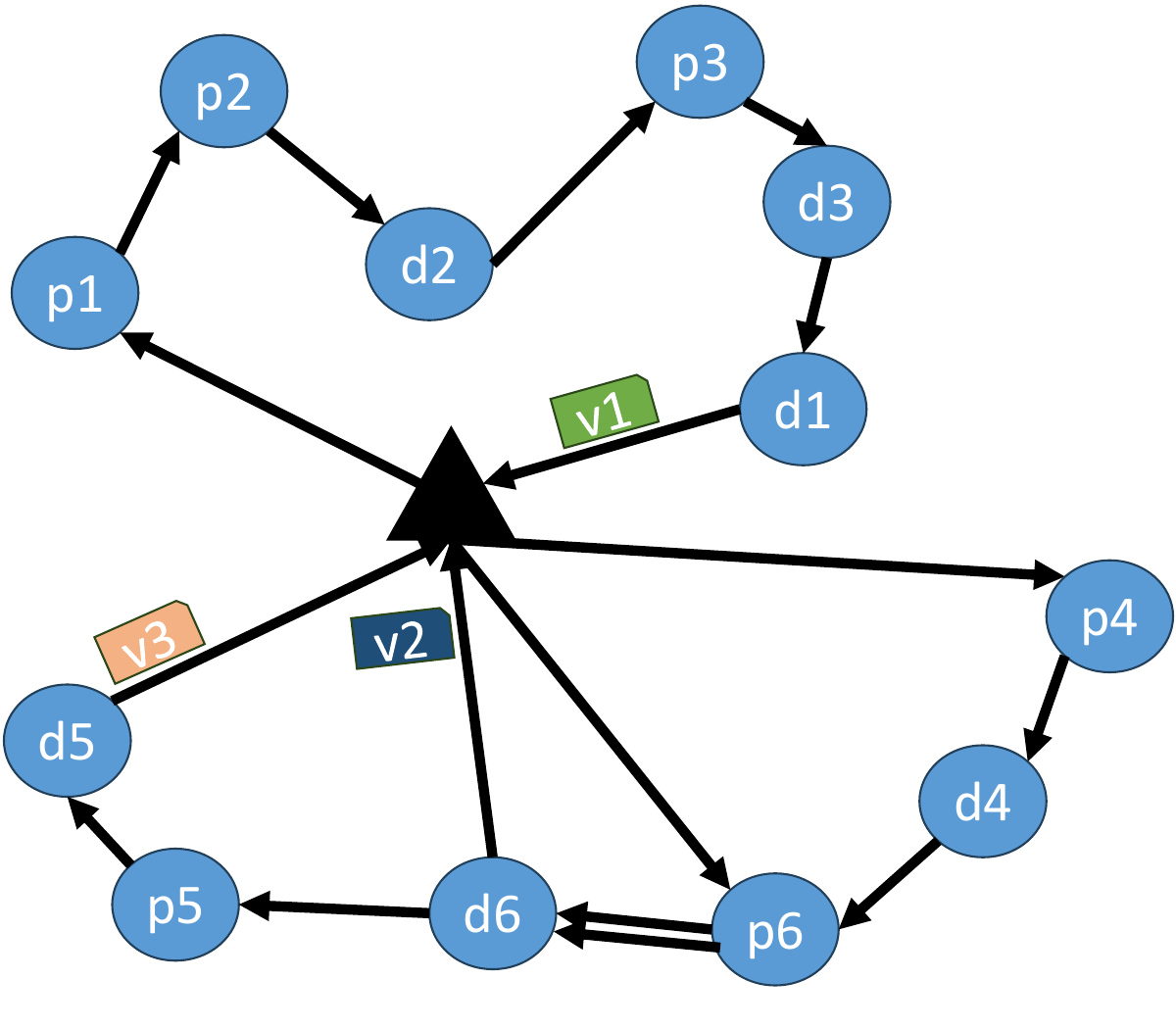}
	\end{minipage}
	\caption{A feasible solution to a DARP-SV instance, depicting its delivery progress at two different times}
\label{fig:DARPSVshow}
\end{figure}

	These developments motivate the Dial-a-Ride Problem with Synchronized Visits (DARP-SV). 
	In DARP-SV, a fleet of vehicles is routed and scheduled to serve customers whose demand may exceed individual vehicle capacity. Large customers require simultaneous service by multiple vehicles, while small customers can be served individually. 
	The problem can also be viewed as a variant of the vehicle routing problem that incorporates synchronization, with the objective of minimizing total routing cost while enforcing synchronized pickups and deliveries when needed. Fig.~\ref{fig:DARPSVshow} illustrates a feasible solution for a DARP-SV instance. Here, $p_i$ and $d_i$ denote the pickup and delivery location of customer~$i$;	$v_1$, $v_2$, and $v_3$ represent three vehicles; and 
    arcs represent vehicle movement. Customer~6 requires joint service by two vehicles, while all other customers have single-vehicle requests.Vehicle $v_1$ sequentially serves customers~1, 2, and 3 independently, while vehicles $v_2$ and $v_3$ must synchronize to simultaneously serve customer~6 (together). This highlights the core feature of DARP-SV: small customers can be handled by a single vehicle, whereas large customers require coordinated service by multiple vehicles.

\vspace{-6mm}
\noindent\scriptsize\textbf{Sources:} \\
\url{https://research.gatech.edu/control-system-helps-several-drones-team-deliver-heavy-packages};\\
\url{https://spectrumnews1.com/.../solo-single-seat-electric-vehicle};\\
\url{https://www.sustainable-bus.com/news/next-modular-vehicles-bus-investment/}.

\normalsize

Our work makes the following four contributions. First, we introduce a new variant of DARP that incorporates synchronization. 
Second, we propose event-based, time-space event-based, and time-space fragment-based formulations for DARP-SV,  we highlight their advantages over arc-based formulations,  and we analyze their relative strengths. An event is defined as a tuple consisting of a location and a set of onboard customers, while a fragment represents a partial path. For DARP, the relative merits of fragment-based and event-based formulations are well established \citep{Rist2021, Gaul2024}. However, their comparative performance in the synchronized setting—where the fragment-based method, unlike the event-based one, now requires a time-space network—remains unexplored.
In this comparison, we also investigate the effects of time-window tightness and request density (customers per unit time). 
Third, we integrate the time-space formulations with  Dynamic Discretization Discovery (DDD)  \citep{Boland2017}, and  show that fragment-based formulations are particularly well suited for combination with DDD (more than arc- and event-based methods). 
Finally, we also apply the fragment-based method with DDD to the classic DARP (without synchronization) and find that, under high request intensity, DDD outperforms the traditional infeasible-constraint callback mechanism \citep{Alyasiry2019, Rist2021}. We further observe that DDD is more effective for DARP than for the pickup-and-delivery problem with time windows (PDPTW).
Pickup-and-delivery problems plan the transportation of items (typically freight) from pickup to delivery nodes at minimal cost \citep{doi:10.1287/trsc.29.1.17}; DARP extends PDPTW by adding customer-oriented constraints, particularly the maximum ride time \citep{Laporte2007}, and is usually applied for transportation of people.

From a practical perspective, this work addresses emerging logistics and passenger transport scenarios where multiple vehicles must be synchronized to serve large customers. Such situations are becoming increasingly relevant as lightweight drones and single- or two-seat modular vehicles shift demand toward smaller-capacity fleets, thereby encouraging coordinated service. From an academic perspective, this study extends the classical DARP to incorporate synchronization into routing decisions. The extended fragment- and event-based formulations (and their time-space counterparts) offer richer comparative insights, clarifying their respective strengths and suitable application settings. Moreover, integrating DDD with fragment-based formulations broadens its applicability beyond arc-based models, yielding faster convergence (requiring fewer iterations).

The remainder of this study is organized as follows. Section~\ref{ch4:liter} reviews the related literature. Section~\ref{DARPSV} formally defines DARP-SV and presents the event-based, time-space event-based, and time-space fragment-based formulations. Section~\ref{DDD} introduces the
DDD algorithm and its adaptation to our problem. Section~\ref{ch4:numericaldiscussion} reports and discusses the computational results. Finally, Section~\ref{ch4:conclusion} concludes the study.

\section{Literature review}	\label{ch4:liter}

We survey three strands of research, namely for
PDPTW, for synchronisation in vehicle routing, and for DDD\@.

\subsection{Pickup and delivery problem with time windows}\label{literPDP}

Traditional exact approaches for PDPTW and DARP include branch-and-cut \citep{doi:10.1287/opre.1060.0283} and branch-and-price (B\&P) \citep{doi:10.1287/trsc.1090.0272,doi:10.1287/opre.1100.0881}. Recent developments, however, show that fragment-based and event-based formulations can provide superior computational performance.

\citet{Alyasiry2019} study PDPTW with a last-in-first-out loading rule, proposing a time-space fragment-based formulation in which callbacks handle rare temporal infeasibilities. A fragment is defined as a partial path where the vehicle is empty only at the first and last node. Building on this idea, \citet{Rist2021} investigate DARP using restricted fragments, each representing a part of a (full) fragment. Their fragment-based formulation, combined with callbacks for time-related constraints, achieves strong performance even without a time-space network. Typically, fragments require more extensive enumeration effort than restricted fragments but capture richer routing information. In many DARP variants, fragments are particularly effective under tight time windows and/or high request intensity, where enumeration remains manageable. \citet{Su2023}, for example, solve an electric autonomous DARP using B\&P with fragments as the basis for path generation, focusing on benchmark instances with tight time windows. Further extensions even employ extended fragments (i.e., fragments augmented with adjacent arcs), which increase enumeration effort but remain tractable for instances with tight temporal structures and/or high request density \citep{Rist2022,Zhang2022}.

A complementary line of work is the event-based formulation \citep{Gaul2024}. Similar to the classical arc-based model of \citet{doi:10.1287/opre.1060.0283}, it represents events as nodes and feasible transitions as arcs. This structure embeds capacity, precedence, and pairing constraints implicitly, reducing the number of explicit constraints and improving solution efficiency. However, as noted by \citet{Gaul2024}, events convey less routing information than fragments. The computational results in \citet{Rist2021} and \citet{Gaul2024} suggest that fragment-based methods tend to perform better on large instances, even though the two papers do not include a direct comparison.



\subsection{Vehicle routing with synchronization}\label{literVRPS}

Prior studies on vehicle routing have examined scenarios where certain customers require multiple vehicles to arrive simultaneously or in a specified sequence. The most relevant category is extra-route synchronization, which coordinates vehicle visits across different routes \citep{drexl2012,Soares2024}. Representative cases include the manpower routing problem \citep{Luo2016} and synchronized delivery \citep{Li2020}. Both involve multiple vehicles or personnel serving the same customer simultaneously and are typically solved using B\&P with branching on time windows. Other related work addresses inter-echelon coordination \citep{Escobar-Vargas2024} and truck-drone synchronization \citep{Dukkanci2024}, where synchronization occurs at designated stations rather than at customer sites. In contrast to these studies, our work not only requires the same vehicles to reach customer sites simultaneously, but also integrates this setting with dial-a-ride.  Existing studies that combine DARP with synchronization primarily focus on passenger transfers \citep{Masson2014,Gkiotsalitis2023,Fu2023} and synchronized  arrivals aligned with school schedules \citep{Vercraene2023};  all these problems have only been addressed using heuristic approaches.

\subsection{Dynamic discretization discovery}\label{literDDD}

Dynamic Discretization Discovery (DDD) \citep{Boland2017} is an iterative refinement method that solves continuous-time problems over a partial time-space network and progressively inserts time indices until optimality. Beyond its original use in service network design and related applications \citep{doi:10.1287/ijoc.2023.0061}, DDD has been successfully applied to a variety of routing problems.  These include cases with extra-route synchronization, such as two-echelon routing \citep{Escobar-Vargas2024} and collision-free multiagent pathfinding via DDD-inspired methods \citep{Adamo2025MAPF_Code}, as well as cases without extra-route synchronization, including the time-dependent shortest path problem \citep{doi:10.1287/ijoc.2021.1084} and vehicle scheduling with limited departure-time flexibility \citep{vanLieshout2025DDDMDVSPTS}. The work most related to ours is that by \citet{sippel2024}, who integrate fragments, DDD, and variable fixing techniques for the PDPTW and show that fragments accelerate DDD convergence. We exploit similar synergies, but in a fundamentally different setting that requires synchronized service at customer nodes. Unlike  \citeauthor{sippel2024}'s PDPTW setting, where all paths have independent time schedules and DDD is merely an optional substitute for callbacks, our DARP-SV requires synchronized time schedules across vehicles, making DDD essential rather than optional in reaching continous-time solutions. Our study of classical DARP in Section~\ref{ch43} will also indicate that fragment-DDD integration behaves differently there than in PDPTW.

\section{Problem statement and linear formulations}\label{DARPSV}

DARP-SV seeks to design routes and schedules for a combination of small and large customers, where large customers require vehicle synchronization. This section first presents a problem statement (Section~\ref{ch4:problemDef}), followed by three linear formulations: an event-based formulation (Section~\ref{ch4:EBF}), a time-space event-based formulation (Section~\ref{ch4:TSEBF}), and a time-space fragment-based formulation (Section~\ref{ch4:fragmentformulate}).

For DARP, fragment-based methods on physical networks and event-based methods have been proposed, so their differences are well understood. For DARP-SV, however, fragment-based methods require a time-space network to model synchronization and a tailored recovery procedure (Section~\ref{DDD}) to reconstruct continuous-time schedules: including continuous-time variables without a time-space network would break the network flow structure and is therefore not considered. In contrast,  event-based methods offer more flexibility: they can either retain the original event-based formulation of \citet{Gaul2024} (already including continuous-time variables), with extensions for synchronization constraints, or adopt a time-space representation. The  fragment-based  time-space formulation builds on \citet{Alyasiry2019} by introducing parameters that allow multiple vehicles to traverse the same fragment and by redefining node arc variables from binary to integer to capture synchronized visits.  Here, a node arc represents the connection between fragments. We compare these models with an arc-based formulation (ABF) adapted from \citet{doi:10.1287/opre.1060.0283}  with minor modifications to incorporate synchronization. The full ABF formulation is provided in Appendix~\ref{ABFappendix}. 


\subsection{Problem description}\label{ch4:problemDef}

DARP-SV involves $n$ customers, with each customer $i$ having a unique pickup and delivery location pair represented as $(i, i+n)$, where $n$ is the total number of customers. The pickup and delivery locations are gathered in the sets $P=\{1,2,\dots,n\}$ and $D=\{n+1,n+2,\dots,2n\}$, respectively. 
The location set $N=P\cup D \cup \{0, 2n+1\}$ consists of all pickup and delivery locations, along with the origin depot location $0$ and the destination depot location $2n+1$. Each location~$i\in P\cup D$ has a specific time window $(e_i, l_i)$, defining the earliest and latest time a vehicle can arrive at that location. The arc set is denoted by~$A$, including all location arcs  $(i,j)\in N \times N$ with $i\ne j$, and the cost and the minimum time associated with traversing the location arc $(i, j)\in A$ are represented by $C_{ij}$ and $T_{ij}$, respectively. Following \citet{Rist2021}, the service time at location~$i$ is embedded in $T_{ij}$ and in the maximum ride time $R_i$, and is therefore not explicitly mentioned. The set of vehicles is represented by~$V$. 
The load change at location $i \in N$ is expressed as~$q_i$, with $q_0 = q_{2n+1} = 0$ and $q_i = -q_{n+i}$ ($q_i \in \mathbb{N}^+$) for $i \in P$. This quantity represents the demand of customer~$i$. At pickup location $i \in P$, the vehicle adds load $q_i$, while at delivery location $i+n \in D$, the vehicle reduces load by $q_i$. The parameter~$Q$ represents vehicle capacity.
Customers are categorized by demand size as either small or large: a small customer~$i$ has a load~$q_i \leq Q$, while a large customer~$j$ has a load~$q_j > Q$. For the former type, we represent the pickup and delivery location sets with $P_s = \{1, 2, \dots, m\}$ and $D_s = \{n+1, n+2, \dots, n+m\}$, respectively, where $m$ represents the total number of small 
customers. For large customers, the pickup and delivery locations are collected in the sets $P_l = \{m+1, \dots, n\}$ and $D_l = \{m+n+1, \dots, 2n\}$, respectively.

The objective of DARP-SV is to minimize travel costs while ensuring that at most $|V|$ vehicles are utilized to visit all pickup and delivery locations, starting from the origin depot location $0$ and ending at the destination depot location~$2n+1$. To achieve this, each pickup location $i\in P$ must be visited by the same vehicle as its corresponding delivery location~$i+n\in D$, with the pickup location visited before the delivery location (precedence). Moreover, each customer~$i\in P$ has a maximum ride time, each location~$i\in N$ has its time window, and each vehicle is limited to a capacity of at most $Q$. 

Our problem statement entails several extra assumptions. First, we assume that each large customer's demand must be transported simultaneously by the corresponding vehicles. We also assume that if a vehicle visits the pickup location of large customer~$i\in P_l$, its load before this location must be empty, and its next location must be $i+n$. Therefore, for customer~$i\in P_l$, $\lceil q_i/Q \rceil$ empty vehicles need to reach its pickup and delivery location at the same time. During network construction, we connect each pickup location of a large customer only to its corresponding delivery location and not to any other node, and arcs from any customer pickup (both small and large) to a large customer pickup are not included. This applies to all formulations of the problem. Second, we assume that all vehicles share identical arc costs and do not account for weight-related costs. If needed, however,  weight-related costs can easily be incorporated into the event-based and fragment-based formulations by adjusting the cost parameters. Third, we do not consider maximum ride time constraints on vehicles (i.e., limits on the total working time of a driver, as opposed to customer ride time), since our primary focus is on the synchronization. Violations of this constraint are expected to be rare and, if necessary, it can be incorporated as a lazy constraint through callbacks.


\subsection{Event-based formulation for DARP-SV}\label{ch4:EBF}

In this subsection, we present the \textit{event-based formulation} (EBF), adapted from the original DARP formulation of \citet{Gaul2024} to suit DARP-SV\@. The main modification introduces vehicle synchronization and two sets of decision variables, namely one indicating whether an event arc is used and the other specifying the number of vehicles traversing it, whereas previous studies use only the former.

	EBF builds on an \textit{event-based network} $G_E=(V_E,A_E)$ with event set $V_E$ and event arc set $A_E$. 
	Each \textit{event} is a tuple $u=(i,S)$, where $i\in N$ is a location (pickup, delivery, or depot) and $S\subseteq P$ is the set of onboard customers immediately after serving~$i$. For simplicity, the currently served pickup is not included in $S$.
	For capacity feasibility, $\sum_{j \in S \cup \{i\}} q_j \le Q$ if $i\in P$, and $\sum_{j \in S} q_j \le Q$ if $i\in D$. 
	An \textit{event arc} connects two events $(i,S),(j,S')\in V_E$, and can be written as $((i,S),(j,S'))$. Here, if $i,j\!\in\!P$ then $S\cup\{i\}=S'$; if $i,j\!\in\!D$ then $S=S'\cup\{j-n\}$; if $i\!\in\!P$ and $j\!\in\!D$ then $S\cup\{i\}=S'\cup\{j-n\}$; and if $i\!\in\!D$ and $j\!\in\!P$ then $S=S'$. Two additional nodes $O^+$ and $O^-$ also belong to $V_E$; the origin depot $O^+=(0,\emptyset)$ is connected only to events~$(i,\emptyset)$ for $i\in P$, and only events~$(i,\emptyset)$ for $i\in D$ are connected to the destination depot $O^-=(2n+1,\emptyset)$.

\begin{figure}[h]
	\centering
		\scalebox{1}[0.9]{
			\begin{tikzpicture}[>=latex, every node/.style={font=\small}]
				\tikzset{
					node/.style={draw, ellipse, inner sep=2pt},
					base/.style={semithick, ->},
				}
				
				\node[node] (Oplus) at (2,0) {$O^+$};
				\node[node] (p2) at (4,0) {$p2$};
				\node[node] (p3set) at (6,0) {$(p3,\{2\})$};
				\node[node] (p1) at (6.5,1.2) {$(p1,\{2,3\})$};     
				
				\node[node] (p3only) at (5.2,-1.0) {$p3$};
				
				\node[node] (d32) at (9.5,1.2) {$(d3,\{1,2\})$};
				\node[node] (d1) at (9.5,-0.5) {$(d1,\{2\})$};
				\node[node] (d3) at (11,0.3) {$d3$};
				\node[node] (d2set) at (6.5,-1.6) {$(d2,\{3\})$};   
				\node[node] (d2) at (9.5,-1.5) {$d2$};
				\node[node] (Ominus) at (13,0) {$O^-$};
				
				\draw[base] (Oplus) -- (p2);
				\draw[base] (p2) -- (p3set);
				\draw[base] (p3set) -- (p1);
				\draw[base] (p1) -- (d32);
				\draw[base] (d32) -- (d1);
				\draw[base] (d1) -- (d2);
				
				\draw[base] (p3set) -- (d2set);
				\draw[base] (d2set) .. controls (8.3,0.6) and (10.0,0.5) .. (d3);
				\draw[base] (d2) -- (Ominus);
				\draw[base] (d3) -- (Ominus);
				
				\draw[base] (p2) -- (d2);
				\draw[base] (Oplus) -- (p3only);
				\draw[base] (d2) .. controls (6,-1) .. (p3only);
				\draw[base] (p3only) .. controls (7.5,-0.4) and (9.8,0.1) .. (d3);
				\draw[base] (d3) .. controls (10,0.8) and (6.5,0.9) .. (p2);
				
			\end{tikzpicture}
		}
		\caption{An event-based network for DARP}
		\label{fig:event_network}
	\end{figure}
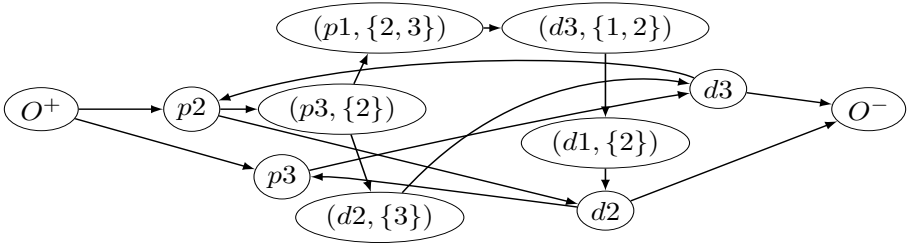
	
The event-based network inherently enforces pairing, precedence, and capacity, as the event representation continuously tracks vehicle loads and event arcs implicitly handle pairing and precedence constraints throughout the network. Events and their arcs are generated by enumeration; for details we refer to \citet{Gaul2024}. Fig.~\ref{fig:event_network} provides an example network. An event is identified as $(p_i,S)$ or $(d_i,S)$, where $p_i$ and $d_i$ denote the pickup and delivery location of customer~$i$, respectively, with $p_i = i$ and $d_i = i + n$, and $S$ is the set of onboard customers. If $S=\emptyset$, the event can be written by its first element. For instance, $(p_1,\{2,3\})$ represents the state after picking up customer~1, when customers~2 and~3 are already on board; $d_2$ represents the state after delivering customer~2 with an empty vehicle.
An example path $(p_2, p_3, p_1, d_3, d_1, d_2)$ then corresponds to the following sequence of feasible events: $p_2$, $(p_3,\{2\})$, $(p_1,\{2,3\})$, $(d_3,\{1,2\})$, $(d_1,\{2\})$, and $d_2$.

	\begin{table}[h]
		\footnotesize
		\centering
		\caption{Decision variables of EBF}
		\begin{tabular}{@{}ll@{}}
			\toprule
			Variables & Definition \\\midrule
			$x_{uv}$ & number of vehicles traversing event arc $(u,v)\in A_E$, $x_{uv}\in\{0,1,\dots,U_{uv}\}$ \\
			$y_{uv}$ & = 1 if the event arc $(u,v)$ is traversed, = 0 otherwise \\
			$t_{i}$  & departure time at customer location $i\in N$ \\
			\bottomrule
		\end{tabular}
		\label{table:ch4notation3.24}
	\end{table}
    
The decision variables of the event-based formulation are summarized in Table~\ref{table:ch4notation3.24}. 
The value~$U_{uv}$ is an auxiliary parameter that equals the maximum number of vehicles that can traverse the event arc $(u,v)\in A_E$.  With $u=(i,S)$ and $v=(j,S')$, it is calculated as $U_{uv} = \min(\lceil |q_i/Q| \rceil, \lceil |q_j/Q| \rceil)$ for $i,j \in P\cup D$, and $U_{O^+v} = \lceil |q_j/Q| \rceil$ and $U_{uO^-} = \lceil |q_i/Q| \rceil$.
The quantity  $c_{uv}= C_{ij}$ denotes the travel cost associated with event arc $(u,v)$.
	The set $A_E(i,j)$ contains all the  event arcs whose location arcs correspond to $(i,j)$, while $A^+(i)$ represents the set of event arcs originating from location~$i\in N$.  Finally,
	$A_E^+(v)$ and $A_E^-(v)$ are the set of event arcs originating from and ending at event~$v\in V_E$, respectively. 
	
	The formulation is as follows:\allowdisplaybreaks
	{\small
		\setlength{\abovedisplayskip}{5pt}
		\setlength{\belowdisplayskip}{5pt}
		\setlength{\jot}{2pt}
		\allowdisplaybreaks
		\begin{align}
			\min \quad & \sum_{(u,v)\in A_E} c_{uv}\, x_{uv}  \label{ch4:EBobj} \\[0.3em]
			\text{s.t.}\quad
			& \sum_{(u,v)\in A_E^-(v)} x_{uv} = \sum_{(v,u)\in A_E^+(v)} x_{vu},
			&& \forall v\in V_E\setminus\{O^+,O^-\} \label{ch4:EBflow} \\
			& y_{uv} \le x_{uv} \le U_{uv}\, y_{uv}, 
			&& \forall (u,v)\in A_E \label{ch4:EByX} \\
			& \sum_{(u,v)\in A^+(i)} x_{uv} = \lceil q_i/Q \rceil,
			&& \forall i\in P \label{ch4:EBpickupcover} \\
			& \sum_{(u,v)\in A^+(0)} x_{uv} \le |V|, \label{ch4:EBdepotstart} \\
			& t_i + T_{ij} - M_{ij}\Bigl(1-\!\!\sum_{(u,v)\in A_E(i,j)} y_{uv}\Bigr)
			\le t_j,
			&& \forall (i,j)\in A \label{ch4:EBtime} \\
			& t_{i+n}-t_i \le R_i,
			&& \forall i\in P \label{ch4:EBride} \\
			& e_i \le t_i \le l_i,
			&& \forall i\in N \label{ch4:EBtw} \\
			& x_{uv}\in \mathbb{N},\; y_{uv}\in\{0,1\},\;		&& \forall (u,v)\in A_E
			\label{ch4:EBdomain1xy}\\
			& t_i\in \mathbb{R}_+ && \forall i\in N 
			\label{ch4:EBdomain1t}
		\end{align}
	}

	The objective function~(\ref{ch4:EBobj}) minimizes the total travel cost. 
	Constraints~(\ref{ch4:EBflow}) enforce flow conservation at events, while~(\ref{ch4:EByX}) link integer flows with binary indicators for the event arcs. 
	Constraints~(\ref{ch4:EBpickupcover}) ensure that each pickup location is served by the required number of vehicles. In the event-based network, the corresponding delivery locations are implicitly visited by the same number of vehicles: for small customers, pickup-delivery pairing is enforced by tracking the onboard customers; for large customers, each pickup is directly connected to its delivery. Constraint~(\ref{ch4:EBdepotstart}) limits the number of vehicles dispatched from the origin depot. 
	Constraints~(\ref{ch4:EBtime}) enforce the time increment, where $M_{ij} =$ \mbox{$\max(0, l_i + T_{ij} - e_j)$}. Because each location~$i\in N$ has a unique departure time~$t_i$, multiple vehicles assigned to the same location must share this value, which ensures synchronization. 
	Constraints~\mbox{(\ref{ch4:EBride})-(\ref{ch4:EBtw})} impose maximum ride time and time window limits. 
	Finally, constraints~(\ref{ch4:EBdomain1xy}) and (\ref{ch4:EBdomain1t}) define the domains of $x$, $y$, and $t$. 
	
	The number of events grows exponentially with vehicle capacity. Consequently, this type of model is well suited to delivery scenarios involving drones, where capacity is typically limited, or other low-capacity vehicles.

	\subsection{Time-space event-based formulation for DARP-SV}\label{ch4:TSEBF}
	
	The time-space event-based formulation (TSEF) extends the EBF by constructing a \textit{time-space event-based network} $G_{TS}=(V_{TS},A_{TE})$, where $V_{TS}$ and $A_{TE}$ denote the set of time-space events and of time-expanded event arcs, respectively. Each time-space event $(v,t)$ combines an event $v=(i,S)\in V_E$ with a discrete time $t\in\mathcal{T}$, where $\mathcal{T}$ is a sufficiently fine set of discretized time values. $A_{TE}$ consists of time-space event arcs and idle event arcs. A time-space event arc $a=(((i,S),t_s),((i',S'),t_e))$ represents the physical movement from $i$ to $i'$ with onboard customers changing from $S$ to $S'$, and the corresponding time index advancing from $t_s$ to $t_e$, where $t_e = t_s + T_{ii'}$ and the travel cost is $C_{ii'}$. Idle event arcs represent waiting between consecutive time indices at zero cost. The travel cost on time-expanded arc $a\in A_{TE}$ is denoted by $\mathcal{C}_a$. The sets $V_{TS}$ and $A_{TE}$ are obtained by combining events and event arcs with their feasible time indices $t\in\mathcal{T}$. The time-space origin and destination depots $tsO^+=(O^+,t_{\min})$ and $tsO^-=(O^-,t_{\max})$ are also included in $V_{TS}$, where $t_{\min}$ and $t_{\max}$ are the minimum and maximum time index in $\mathcal{T}$, respectively.
	
	\begin{table}[h]
		\footnotesize
		\centering
		\caption{Decision variables of TSEF}
		\begin{tabular}{@{}ll@{}}
			\toprule
			Variables & Definition \\\midrule
			$\chi_a$ & number of vehicles traversing time-expanded event arc $a\in A_{TE}$, $\chi_a\in\{0,1,\dots,U_a\}$ \\
			$\gamma_a$ & = 1 if time-expanded event arc $a\in A_{TE}$ is traversed, = 0 otherwise \\
			\bottomrule
		\end{tabular}
		\label{table:TSEBvar}
	\end{table}
	
	The decision variables for the TSEF are summarized in Table~\ref{table:TSEBvar}. The parameters $U$ are computed similarly as in Section~\ref{ch4:EBF}.
	The sets $A_i^{\mathrm{TE}+}$ and $A_i^{\mathrm{TE}-}$ contain all time-expanded event arcs 
	originating from and ending at location~$i\in N$, respectively, while  
	$A_{TE}^+(v,t)$ and $A_{TE}^-(v,t)$ denote the set of time-expanded event arcs 
	originating from and ending at time-space event $(v,t)\in V_{TS}$, respectively. Finally, $s^{(a)}$ and $e^{(a)}$ represent the (known) departure and arrival time of each time-space event arc $a$.

	The formulation can now be stated as follows:
	{\small
		\setlength{\abovedisplayskip}{5pt}
		\setlength{\belowdisplayskip}{5pt}
		\setlength{\jot}{2pt}
		\allowdisplaybreaks
		\begin{align}
			\min \quad & \sum_{a\in A_{TE}} \mathcal{C}_a\, \chi_a \label{TSEBobj}\\[0.3em]
			\text{s.t.}\quad
			& \sum_{a\in A_{TE}^-(v,t)} \chi_a = 
			\sum_{a\in A_{TE}^+(v,t)} \chi_a,
			&& \forall (v,t)\in V_{TS}\setminus\{tsO^+,tsO^-\} \label{TSEBflow}\\
			& \gamma_a \le \chi_a \le U_a\, \gamma_a,
			&& \forall a\in A_{TE} \label{TSEBlink}\\
			& \sum_{a\in A_i^{\mathrm{TE}+}} \chi_a =  \lceil q_i/Q \rceil,
			&& \forall i\in P \label{TSEBcoverpick}\\
			& \sum_{a\in A_{0}^{\mathrm{TE}+}} \chi_a \le |V|, \label{TSEBZ}\\
			& \sum_{a\in A_{i+n}^{\mathrm{TE}-}} s^{(a)} \gamma_a
			-\sum_{a\in A_i^{\mathrm{TE}+}} e^{(a)} \gamma_a 
			\le  R_i,
			&& \forall i\in P \label{TSEBride}\\
			& \chi_a\in \mathbb{Z}_{+},\; \gamma_a\in\{0,1\},
			&& \forall a\in A_{TE} \label{TSEBdomain}
		\end{align}
	}

	The objective function~\eqref{TSEBobj} minimizes the total travel cost.  
	Constraints~\eqref{TSEBflow} ensure flow conservation at each time-space event.  
	Constraints~\eqref{TSEBlink} link integer flows $\chi_a$ with binary arc usage $\gamma_a$, 
	ensuring that a positive flow can exist only if the corresponding event arc is activated.  
	Constraints~\eqref{TSEBcoverpick} guarantee that each pickup is served by the required number of vehicles, 
	while constraint~\eqref{TSEBZ} limits the number of vehicles dispatched from the depot to at most $|V|$.  
	Constraints~\eqref{TSEBride} enforce the maximum ride time for each customer $i \in P$ by bounding the difference between the arrival time at the delivery location and the departure time at the corresponding pickup location of the assigned vehicle. 
	These constraints are exact 
    only when the time discretization is sufficiently fine; 
    otherwise, the ride time limit may not be accurately satisfied or may become overconstrained. Concretely, in TSEF the ride-time constraints are enforced only at discrete time points, which may introduce approximation errors. In contrast, the fragment-based formulation in Section~\ref{ch4:fragmentformulate} embeds ride-time limits within fragments, thereby avoiding this discretization-induced inaccuracy. Finally, constraints~\eqref{TSEBdomain} define the integrality and binary nature of the decision variables.

	\subsection{Time-space fragment-based formulation for DARP-SV}\label{ch4:fragmentformulate}

\subsubsection{General observations}

In the standard DARP, the event-based formulation is structurally similar to arc-based formulations and relies on continuous-time variables combined with big-$M$ constraints to model temporal feasibility. Fragment-based formulations, in contrast, avoid continuous-time variables by operating on a network structure that implicitly enforces temporal constraints, thereby preserving a strong network flow structure and avoiding big-$M$ constraints, which largely explains their computational efficiency. In the work of \citet{Rist2021}, time-space networks are even omitted for fragment-based DARP formulations, as temporal infeasibilities are rare under relaxed time windows (dependent on the benchmark instances) and can be handled efficiently via infeasible-path callbacks, where each path has an independent time schedule.

In the context of DARP-SV, synchronization constraints fundamentally change this setting. While the event-based formulation already contains continuous-time variables and can accommodate synchronization with relatively minor extensions, fragment-based formulations can no longer rely on independent path schedules. Synchronization introduces explicit time-dimensional coupling across vehicles, which invalidates the callback-based approach and necessitates a time-space network representation. As a result, for DARP-SV, the relevant comparison is no longer between event-based formulations and fragment-based formulations on a physical network with callbacks (as in classical DARP), but rather between event-based formulations and time-space fragment-based formulations.

	Building on \citet{Alyasiry2019} and \citet{Rist2021}, we develop the time-space fragment-based formulation (TSFrag) for DARP-SV by introducing fragment parameters that allow multiple vehicles to traverse the same fragment and redefining the node arc decision variable from binary to integer to enable synchronized visits for large customers. In the following, we first introduce the (physical) fragment-based network for DARP, then extend it to the time-space fragment-based network for DARP-SV, and finally present TSFrag.

	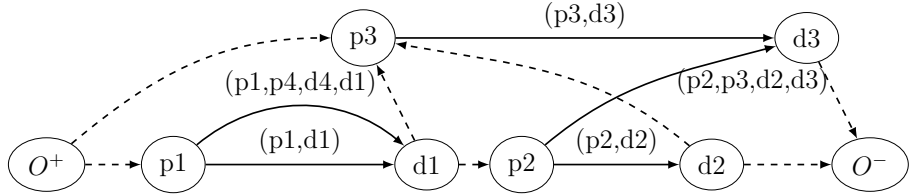
\begin{figure}[H]
		\centering
		\resizebox{0.8\textwidth}{!}{
			\begin{tikzpicture}
				\node[draw,ellipse] (0+) at (-8,1) {$O^+$};
				\node[draw, ellipse] (0-) at (5,1) {$O^-$};
				\node[draw, ellipse] (p2) at (-0.5,1) {p2};
				\node[draw, ellipse] (p3) at (-3,3) {p3};
				\node[draw, ellipse] (p1) at (-6,1) {p1};
				\node[draw, ellipse] (d2) at (2.5,1) {d2};
				\node[draw, ellipse] (d3) at (4,3) {d3};
				\node[draw, ellipse] (d1) at (-2,1) {d1};
				
				\draw[line width=0.75pt, ->,dashed,>=latex] (0+) to[out=40, in=-180]  (p3);
				\draw[line width=0.75pt, ->,dashed,>=latex] (0+) ->  (p1);
				\draw[line width=0.75pt, ->,dashed,>=latex] (d2) ->  (0-);
				\draw[line width=0.75pt, ->,dashed,>=latex] (d3) ->  (0-);
				\draw[line width=0.75pt, ->,dashed,>=latex] (d2)  to[out=140, in=-10]  (p3);
				\draw[line width=0.75pt, ->,dashed,>=latex] (d1) -> (p3);
				\draw[line width=0.75pt, ->,dashed,>=latex] (d1) -> (p2);
				
				\draw[line width=0.75pt, ->,>=latex] (p2) -- node[above] {(p2,d2)} (d2);
				\draw[line width=0.75pt, ->,>=latex] (p3) -- node[above] {(p3,d3)} (d3);
				\draw[line width=0.75pt, ->,>=latex] (p1) -- node[above] {(p1,d1)} (d1);
				\draw[line width=0.75pt, ->,>=latex] (p1) to[out=40, in=140] node[above] {(p1,p4,d4,d1)} (d1);
				\draw[line width=0.75pt, ->,>=latex] (p2)  to[out=40, in=-165] node[right=6pt]{(p2,p3,d2,d3)} (d3);
		\end{tikzpicture}}
		\caption{A fragment-based network for DARP}\label{ch4:fig3}
	\end{figure}
	

\subsubsection{The  networks}

We begin by introducing the fragment-based network. Following \citet{Rist2021}, a \textit{DARP route} is a sequence $(0,i_1,\dots,i_L,2n+1)$ for which a feasible schedule exists, satisfying pairing, precedence, capacity, maximum ride time, and time window constraints. Any part of the route is referred to as a route path. A \textit{fragment} is a route path that starts at a pickup node and ends at a delivery node, with an empty load at and only at both endpoints. In this study, nodes correspond directly to locations. An example of feasible fragments would be $(p_1,p_2,d_2,d_1)$ and $(p_1,p_2,d_1,p_3,d_2,d_3)$, whereas $(p_1,d_1,p_2,p_3,d_2,d_3)$ is not a fragment because the vehicle becomes empty after~$d_1$. Each fragment also satisfies all DARP constraints. A \textit{node arc} connects a delivery or the origin depot to a pickup, or connects a delivery to the destination depot. A fragment-based network is constructed by enumerating all feasible fragments and linking them through node arcs. A feasible path alternates between node arcs and fragments, starting with a node arc from the origin depot to pickups, continuing with a fragment to deliveries, and then with a node arc to either new pickups or the destination depot. Fig.~\ref{ch4:fig3} illustrates this network structure, where ellipses represent nodes, solid lines denote fragments, and dashed lines are node arcs. Note that very long fragments, such as $(p_1,p_2,d_1,p_3,d_2,\dots,p_{100},d_{99},d_{100})$, are theoretically possible, and so is an extremely large fragment count. However, prior studies have shown that this number remains manageable when time windows and maximum ride time are tight \citep{Rist2022,Zhang2022,Su2024}.

	\begin{figure}[h]
		\centering
		\begin{minipage}[t]{0.49\textwidth}
			\centering
			\resizebox{\textwidth}{!}{ 
				\begin{tikzpicture}[every node/.style={font=\large}]
					\tikzset{
						node/.style={circle, fill=blue, inner sep=2pt, draw=none},
						path/.style={ultra thick},
						label/.style={font=\large},   
						time/.style={font=\large},    
					}
					
					\foreach \x in {0,1,...,70} {
						\draw[gray!30] ({0.1*\x}, 0.5) -- ({0.1*\x}, -7.5);
					}
					\node[time] at (0,1) {$t_0$};
					\node[time] at (1,1) {$t_{10}$};
					\node[time] at (2,1) {$t_{20}$};
					\node[time] at (3,1) {$t_{30}$};
					\node[time] at (4,1) {$t_{40}$};
					\node[time] at (5,1) {$t_{50}$};
					\node[time] at (6,1) {$t_{60}$};
					\node[time] at (7,1) {$t_{70}$};
					
					\foreach \y in {0,1,...,7} {
						\node[label, left] at (-0.5,{-1*\y}) {$L_{\y}$};
					}
					
					\node[circle, fill=black, inner sep=2pt, label=below:{Origin}] (origin) at (0,0) {} node[label, right] {$O^+$};
					\node[circle, fill=black, inner sep=2pt] (destination) at (7,-7) {};
					\node[label, left] at (destination) {$O^-$};
					
					\draw[dashed] (0,-0.05) -- (1.3,-1.05) node[node] (p7){} node[label,above] {$p7$};
					\draw[darkgray, ultra thick] (p7) to[out=-20,in=145] node[midway, above right=-3pt, label] {$p7$-$d7$} (2.4,-2) node[node] (d7){} node[label, right=1pt] {$d7$};
					
					\draw[darkgray, ultra thick] ($(p7)-(0.075,0.075)$) to[out=-20,in=145] ($(d7)-(0.075,0.075)$);
					
					\draw[dashed] (d7) -- (3.4,-3) node[node] (p3){} node[label,above right] {$p3$};
					\draw[darkgray, ultra thick] (p3) to[out=-20,in=145] node[midway, above right=-2pt, label] {$p3$-$d3$} (5,-4) node[node] (d3){} node[label, right] {$d3$};
					
					\draw[dashed] (d7) -- (3.4,-5) node[node] (p2){} node[label,above right] {$p2$};
					\draw[darkgray, ultra thick] (p2) to[out=-20,in=145] node[midway, below left=-1pt, label] {$p2$-$p5$-$d2$-$d5$} (6,-6) node[node] (d5){} node[label,below left] {$d5$};
					
					\draw[dashed] (d3) -- (7,-7);
					\draw[dashed] (d5) -- (7,-7);
				\end{tikzpicture}
			}
			\caption{\scriptsize  Time-space fragment-based network}\label{TSRFBNBasic}
		\end{minipage}
		\hfill
		\begin{minipage}[t]{0.50\textwidth}
			\centering
			\resizebox{\textwidth}{!}{ 
				\begin{tikzpicture}[every node/.style={font=\large}]
					\tikzset{
						node/.style={circle, fill=blue, inner sep=2pt, draw=none},
						rnode/.style={circle, fill=purple, inner sep=2pt, draw=none},
						path/.style={ultra thick},
						label/.style={font=\large},   
						time/.style={font=\large},    
					}
					
					\foreach \x in {0,1,...,6} {
						\draw[gray!30] (\x,0.5) -- (\x,-7.5);
						\node[time] at (\x,0.8) {$t_{\x}$};
						\node[label, left] at (-0.5,-\x) {$L_{\x}$};
					}
					\draw[gray!30] (7,0.5) -- (7,-7.5);
					\node[time] at (7,0.8) {$t_7$};
					\node[label, left] at (-0.5,-7) {$L_{7}$};
					
					\node[circle, fill=black, inner sep=2pt, label=below:{Origin}] (origin) at (0,0) {} node[label, right=1pt] {$O^+$};
					\node[circle, fill=black, inner sep=2pt] (destination) at (7,-7) {};
					\node[label, left=4pt] at (destination) {$O^-$};
					
					\draw[dashed] (origin) -- (1.3,-0.95) node[node] (p7){} node[label,above] {$p7$};
					\node[rnode] (p7r) at (1,-0.995) {}; 
					\node[node] (d7) at (2.4,-2) {};
					\node[rnode] (d7r) at (2.0,-2) {}; 
					\node[label,above right] at (d7) {$d7$};
					\draw[darkgray, ultra thick] (p7r) to[out=-20,in=145] node[midway, above right=-1pt, label] {} (d7);
				\draw[darkgray, ultra thick] ($(p7r)-(0.075,0.075)$) to[out=-20,in=145] ($(d7)-(0.075,0.075)$);
				\draw[dashed] (d7r) -- (3.3,-3) node[node] (p3){}; 
				\node[label,above] at ($(p3)+(0.3,0)$) {$p3$};
				\node[rnode] (p3r) at (3,-3) {};
				\node[node] (d3) at (4.4,-4) {}; 
				\node[rnode] (d3r) at (4.0,-4) {}; 
				\node[label,right] at (d3) {$d3$};
				
				\draw[dashed] (d7r) -- (3.3,-5) node[node] (p2){}; 
				\node[label,above] at ($(p2)+(-0.6,0)$) {$p2$};
				\node[rnode] (p2r) at (3,-5.0) {};
				\node[node] (d5) at (5.4,-6) {};
				\node[rnode] (d5r) at (5.0,-6) {}; 
				\node[label,below left] at (d5) {$d5$};
				
				\draw[dashed] (d3r) -- (destination);
				\draw[dashed] (d5r) -- (destination);
				
				\draw[darkgray, ultra thick] (p3r) to[out=-20,in=145] node[midway, right=1pt, label] {} (d3);
			\draw[darkgray, ultra thick] (p2r) to[out=-20,in=145] node[midway, below left=0.1	pt, label] {} (d5);
		
	\end{tikzpicture}

}
\caption{\scriptsize Partial time-space fragment-based network (DDD, Section~\ref{DDD})}\label{TSRFBNround}
\end{minipage}
\vspace{0.5em} 
\begin{flushleft}
\scriptsize
\textit{Note:} The vertical axis represents space and the horizontal axis represents time. The light gray background lines indicate discretized time indices. A solid line denotes a time-space fragment, while a dashed line indicates a time-expanded arc. The route path associated with each fragment is presented alongside it. Blue and purple nodes that are close to each other correspond to the same spatial locations with actual and rounded time indices, respectively.
\end{flushleft}

\end{figure}


We now construct a \textit{time-space fragment-based network} for DARP-SV, following the same procedure as for the time-space event-based network. The key distinction is that, whereas a departure time $t$ on an event arc uniquely determines the arrival time $t'$ at its endpoint, a fragment can span multiple location arcs and implicitly enforce time window constraints for all locations within it. Therefore, for each fragment, the arrival time $t'$ at the end node is set to the earliest feasible arrival time that satisfies all time window and maximum ride time constraints. 

We denote the time-space fragment-based network by $G(N_N, F, A_N)$, with $N_N$ the set of time-space nodes, $F$ the set of time-space fragments, and $A_N$ containing all time-expanded node arcs.  An illustration can be found in Fig.~\ref{TSRFBNBasic}.
A time-space node $h \in N_N$ is defined as $(p,t)$, where $p$ is a location and $t\in \mathcal{T}$ a feasible time index. The set $N_N$ consists of the time-space depots ($tsO^+, tsO^-$), the time-space pickup nodes, and the time-space delivery nodes. A time-space fragment $f \in F$ is defined as $(h,\text{path}(f),h')$, where $h=(p,t)$ and $h'=(p',t')$ are the start and end nodes, and $\text{path}(f)$ is the (physical) route path. A time-expanded node arc $a \in A_N$ connects two time-space nodes and can be of two types: (i) an idle node arc, representing waiting over time, or (ii) a time-space node arc, representing movement in both time and space. The former has zero cost, whereas the latter incurs the travel cost of the underlying location arc. 

\subsubsection{The formulation} Time-space fragments are categorized as (i) covering the pickup-to-delivery movements of one or more small customers, or (ii) representing the pickup-to-delivery movement of a single large customer. We define $\alpha_f$ as the number of vehicles associated with a time-space fragment or time-expanded node arc $f \in F \cup A_N$. For a time-space fragment, $\alpha_f = 1$ for type~(i) and $\alpha_f = \lceil q_i / Q \rceil$ for type~(ii) involving customer \mbox{$i \in P_l$}. For each time-expanded node arc $a = ((p,t),(p',t')) \in A_N$, $\alpha_a = \lceil \min ( |q_{p}|, |q_{p'}| ) / Q \rceil$. Fig.~\ref{TSRFBNBasic}  illustrates a feasible solution over a time-space fragment-based network. Two vehicles simultaneously traverse the fragment $p7$--$d7$ associated with large customer~$7$, entering $d7$ with a fragment flow of two vehicles and departing via two distinct node arcs. The fragment cost is defined as the sum of the costs of all location arcs along its route path. 

For notation, let $F_h^+$ and $F_h^-$ denote the sets of fragments starting from and ending at node $h \in N_N$, respectively; 
$F_i$ the set of fragments traversing location $i \in P \cup D$; and $A_h^+$ and $A_h^-$ the sets of arcs leaving from and entering node $h \in N_N$, respectively. Value $\kappa_{f}$ denotes the cost associated with each time-space fragment or time-expanded node arc $f \in F \cup A_N$.


The decision variables of TSFrag are presented in Table~\ref{table:ch3notation3.23}.

\begin{table}[h]
\footnotesize
\centering
\caption{Decision variables of TSFrag}
\begin{tabular}{@{}ll@{}}
\toprule
Variables & Definition \\\midrule
$X_{f}$ & = 1, if the time-space fragment  $f\in F$ is traversed; = 0, otherwise \\
\multirow{1}{*}{$Y_{a}$} & the number of vehicles traversing the time-expanded node arc~$a\in A_N$, 
$Y_{a}\in \{0,1,\dots,\alpha_a\}$ \\\bottomrule
\end{tabular}
\label{table:ch3notation3.23}
\end{table}

The TSFrag formulation is presented below. Note that the pairing, precedence, and ride time constraints are already enforced within each fragment.

{\small
\setlength{\abovedisplayskip}{5pt}
\setlength{\belowdisplayskip}{5pt}
\setlength{\jot}{2pt}
\allowdisplaybreaks
\begin{align}
\min\quad 
& \sum_{f\in F} \kappa_f\, \alpha_f X_f + \sum_{a\in A_N} \kappa_a\, Y_a 
\label{ch4:MDARPobj2}\\[0.3em]
\text{s.t.}\quad
& \sum_{f\in F_h^-} \alpha_f X_f + \sum_{a\in A_h^-} Y_a
= \sum_{f\in F_h^+} \alpha_f X_f + \sum_{a\in A_h^+} Y_a,
&& \forall h\in P_N\cup D_N \label{ch4:FragmentArc2}\\
& \sum_{f\in F_i} X_f = 1,
&& \forall i\in P \label{ch4:cover3}\\
& \sum_{a\in A_{tsO^+}^+} Y_a \le |V|, \label{ch4:vehicle2}\\
& X_f \in \{0,1\},
&& \forall f\in F \label{ch4:DARdomain31}\\
& Y_a \in \{0,1,\dots,\alpha_a\},
&& \forall a\in A_N \label{ch4:DARdomain32}
\end{align}
}

The objective function~(\ref{ch4:MDARPobj2}) minimizes the total weighted cost of traversed time-space fragments and time-expanded node arcs. 
Constraints~(\ref{ch4:FragmentArc2}) describe the network flow for the time-space fragments and time-expanded node arcs.
This can be divided into two scenarios: network flow at a time-space pickup node belonging to $P_N$ or at a time-space delivery node belonging to $D_N$, since a time-space fragment can only start at a time-space pickup node and end at a time-space delivery node.
Constraint~(\ref{ch4:cover3}) guarantees that each pickup location is visited once. In addition, given the parameter $\alpha_f$, a fragment may be served by multiple vehicles simultaneously to accommodate large customers.
Constraints~(\ref{ch4:vehicle2}) state that at most $|V|$ vehicles are used. 
Constraints~(\ref{ch4:DARdomain31}) and (\ref{ch4:DARdomain32}) specify the domains of $X$ and~$Y$, respectively.

Subtours may occur when the time discretization is coarse.  As an example, consider the potential subtour $(p_1,p_2,d_1,d_2,p_1)$, where $(p_1,p_2,d_1,d_2)$ forms a fragment and $(d_2,p_1)$ is a node arc.
The actual visiting times for this sequence might be $(\mbox{$10:00$}, \mbox{$10:06$}, \mbox{$10:07$}, \mbox{$10:08$}, \mbox{$10:09$})$.   
If the time discretization unit is five minutes, this schedule would be rounded to the following discretized schedule: $(\mbox{$10:00$}, \mbox{$10:05$},\\ \mbox{$10:05$}, \mbox{$10:05$}, \mbox{$10:05$})$, and the subtour would never be created because of the time incrementation. Under an alternative 10-minute rounding scheme, by contrast, the corresponding discretized schedule would be
$(\mbox{$10:00$}, \mbox{$10:00$}, \mbox{$10:00$}, \mbox{$10:00$}, \mbox{$10:00$})$ and the formulation would allow this subtour. 
Time-space formulations can prevent many subtours because increasing time indices force the start and end nodes of an arc (or event arc / fragment) to be associated with different time indices. As a result, infeasible subtours in the physical network may disappear once embedded in the time-space network. When the time-space network is constructed by rounding down arc travel times, however, some arcs may end up with zero length, and the time-space formulation provides less protection against subtours.
Consequently, the finer the time discretization, the less likely subtours are to appear. Any occurring subtours are eliminated through lazy-constraint callbacks at each incumbent solution.	If a subtour with $c>2$ (physical) fragments $(f_1,\dots,f_c)$ and $c-1$ node arcs $(a_1,\dots,a_{c-1})$ is detected, it is removed by enforcing
$\sum_{k=1}^c \sum_{f \in F^{(f_k)}} X_{f} 
+ \sum_{k=1}^{c-1} \sum_{a \in A_N^{(a_k)}} Y_{a} 
\le 2c - 2$, where $F^{(f_k)}$ and $A_N^{(a_k)}$ are the time-space fragments and time-space node arcs derived from (physical) fragment $f_k$ and node arc $a_k$, respectively.  The same procedure applies to TSEF by replacing fragments and node arcs with event arcs.

The fragment count grows exponentially with vehicle capacities, time window size, and maximum ride times, which may hinder scalability. Nonetheless, for many delivery scenarios with low-capacity vehicles and drones, these parameters are typically small, making the proposed framework well suited.

\section{Dynamic discretization discovery}\label{DDD}

DDD is an iterative algorithm for optimally solving continuous-time service network design problems over partial time-space networks \citep{Boland2017}. In this study, we adapt DDD to TSEF and TSFrag. Since TSFrag requires more adjustments than TSEF and thus the procedure developed for TSFrag can be directly applied to TSEF, we focus primarily on describing the design of DDD for TSFrag.

Let $G(\mathcal{T})$  denote the full time-space network, which is constructed from the (physical) network and the (full) set of time indices $\mathcal{T}$. 
At iteration~$k$, the algorithm operates on a partial network $G(\mathcal{T}^{k})$ using a coarse time index set $\mathcal{T}^{k}\subseteq \mathcal{T}$. 
For simplicity, we generally use the term ``arc” to refer to both fragment and node arcs when distinction is not necessary. In the construction of $G(\mathcal{T}^{k})$, each (physical) arc~$(i,j)$'s actual length (i.e., its minimum travel time $T_{ij}$) is rounded down due to limited time indices in $\mathcal{T}^{k}$, while the cost parameter for each arc remains unchanged. 

DDD contains four steps. 

\noindent \textbf{Step~1} constructs a partial time-space network $G(\mathcal{T}^k)$ with rounded time indices and formulates the relaxed problem $\mathcal{P}_{\mathcal{T}^k}$. The rounding will shorten some of the the arc travel times (see Fig.~\ref{TSRFBNround} and Appendix~\ref{DDDstep1} for details), which relaxes some of the time-related constraints such as time windows and maximum ride times. 

\noindent \textbf{Step~2} solves $\mathcal{P}_{\mathcal{T}^k}$. 
The optimal solution $S(\mathcal{P}_{\mathcal{T}^k})$ provides a valid lower bound for the original problem $\mathcal{P}$. 
Since it is obtained using shortened arc lengths, the corresponding paths (a set of physical paths with synchronization requirements) from $S(\mathcal{P}_{\mathcal{T}^k})$ may not be feasible in $G(\mathcal{T})$  but are guaranteed to be feasible in $G(\mathcal{T}^k)$.

\noindent \textbf{Step~3} checks the feasibility of the paths from $S(\mathcal{P}_{\mathcal{T}^k})$ and, in case of infeasibility, selects the arcs to be lengthened. 
The check is performed by a selection model of \citet{Boland2017} that tests whether a feasible schedule exists with the actual arc lengths. If such a schedule exists, the algorithm terminates since a solution that provides a lower bound and is also feasible in $G(\mathcal{T})$  must be optimal for $\mathcal{P}$. 
Otherwise, to preserve the existence of a feasible schedule for the current solution paths, a subset of arcs must continue to retain shortened arc travel times; the selection model will try to allow as many arcs as possible to use their actual travel time and thus identify the arcs that must keep their shortened arc length to ensure the existence of a feasible schedule.  The identified arcs will then be lengthened to eliminate such infeasibility in future iterations. 

\noindent \textbf{Step~4} adds new time indices based on the identified arcs. For each selected arc $(p, p')$, new arrival times are generated by adding the actual arc time length to each feasible time index~$t$ of $p$. All resulting values $t' = t + T_{pp'}$ that fall within the time window of $p'$ are inserted into its time index set. The algorithm then returns to Step 1 and repeats until reaching optimality. This iterative refinement guarantees convergence to an optimal feasible solution; see \citet{Boland2017} for theoretical analysis and proofs.  It can be verified that the convergence result is also valid for our implementation.

Compared to \citet{Boland2017} and related studies, our use of DDD differs in both problem structure and methodological design. 
On the problem side, DARP-SV requires synchronization directly at customer locations under strict time windows, whereas prior studies typically address synchronization at intermediate facilities or along routes. 
This tighter coupling of vehicle paths increases the difficulty of finding feasible solutions and calls for a careful assessment of DDD's applicability. Methodologically, we apply DDD to fragment-based rather than arc-based formulations, which may lead to different iteration counts. Since event- and arc-based models share the same structure, TSEF with DDD can be viewed as an arc-based formulation with DDD.

Several modifications are needed when applying DDD to TSFrag and TSEF\@. For TSFrag, a key parameter is the shortened arc time length used in the selection model: since each fragment may contain multiple location arcs and only the fragment-level value is directly available, the time length for each location arc is derived as its actual travel time minus the whole fragment's rounding discrepancy. Meanwhile, we constrain each fragment to a single rounding discrepancy (see Appendix~\ref{DDDselctionmodel}). When executing DDD for TSEF, minor inaccuracies may arise from the inherent limitation of constraint~\eqref{ch4:EBride}, where continuous-time restrictions meet discrete-time differences. For example, consider a route $(p_1, p_2, d_1, d_2)$ with feasible times \mbox{$10:00$}, \mbox{$10:24$}, \mbox{$10:26$}, and \mbox{$10:50$}, and a maximum ride time of 26 minutes for both requests. If the time discretization uses 10-minute intervals, then only schedules such as \mbox{$10:00$} -- \mbox{$10:20$} -- \mbox{$10:20$} -- \mbox{$10:50$} (ride times 20 and 30 minutes for customer~1 and 2, respectively) or \mbox{$10:20$} -- \mbox{$10:30$} -- \mbox{$10:30$} -- \mbox{$10:50$} (30 and 20 minutes) are possible, both of which violate the ride time constraint. Then, the feasible path would be eliminated and thus the obtained solution is not a lower bound. Relaxing this ride-time limit might guarantee optimality but would also increase DDD's iterations and computational effort, thereby weakening TSEF's role as a benchmark algorithm. We therefore retain TSEF as an approximate benchmark and disregard this minor discrepancy.

\section{Computational results}\label{ch4:numericaldiscussion}

This section presents the results of our computational experiments. We first evaluate our procedures on DARP-SV, the main focus of this study, and then apply the methods also to classical DARP and PDPTW, which is possible without modifications since DARP-SV is a generalization of both DARP and PDPTW.

\subsection{Instance generation and experimental design}\label{ch41}

We conduct two experiments: the first evaluates our formulations on DARP-SV using two benchmark instance sets, and the second compares them with a state-of-the-art method \citep{Rist2021, Alyasiry2019} for classical DARP and  for PDPTW.

In the first experiment, we assess the computational performance of DARP-SV using two benchmark datasets. The first dataset is derived from the 48 classical DARP instances introduced by \citet{doi:10.1287/opre.1060.0283}, which have been widely used in subsequent studies, including \citet{Ropke2007}, \citet{Gschwind2015}, and \citet{Rist2021}. These instances are either of Type~A (moderately tight windows, unit demands, capacity $=3$) or of  Type~B (wider windows, demands 1 to 6, capacity $=6$), each with 15-minute pickup and delivery windows over an 8-12 hour horizon and a 30-minute ride-time limit. To construct synchronized-service instances, we designate one third of the customers as large ($R_{L}=1/3$), assign each a load equal to twice the vehicle capacity, and triple the fleet size compared to the original instances. 
This defines the instances in  ``Set~1,'' which have normal temporal settings and low request density. 

The second dataset introduces tighter windows and high request intensity, reflecting practical shared-mobility or drone-delivery conditions. Starting from Set~1, we compute feasible time windows using standard reachability checks and compress all pickup times into a single hour; we refer to Appendix~\ref{CompResults:preprocessing} for details of both steps. Pickup-window widths take $P_{TW}\in\{15,30\}$, and delivery windows follow from direct travel times and a maximum-ride-time factor $P_{\mathrm{De}}\in\{1.5,2.0\}$. Synchronization rules remain unchanged. We restrict our analysis to Type-A instances because their temporal differences from Type-B are superseded by our modification. Using a fleet size of four times the original size, we obtain 13 instances that are feasible under all parameter settings. We additionally test the scenario where the fleet size is enlarged by factor five for $R_{L}=1/3$ and $P_{TW}=15$, to complete the comparison. In this experiment, we evaluate four formulations: ABF, EBF, TSEF+DDD, and TSFrag+DDD, where TSEF+DDD and TSFrag+DDD denote TSEF and TSFrag combined with DDD\@. Both DDD-based procedures employ an initial coarse 50-minute resolution, where the choice of 50 minutes simply represents a large initial time step. The time index set of each node additionally includes its earliest departure time and latest arrival time.

In the second experiment, we compare EBF, TSFrag+DDD, and TSFrag+C for classical DARP and PDPTW\@. Here, TSFrag+C uses TSFrag to enforce the main constraints and callbacks for the remaining temporal ones \citep{Rist2021, Alyasiry2019}, but this design is highly sensitive to tight temporal settings, where frequent callbacks substantially increase computational burden. Moreover, this approach cannot be applied to DARP-SV because callbacks operate only on paths with independent schedules. The experiment has two objectives: 
(i) to compare the performance of TSFrag+DDD with TSFrag+C, and (ii) to examine whether DDD provides different benefits in DARP and in PDPTW\@. For DARP, we use the second DARP-SV dataset with $R_{L}=0$, $P_{TW}=15$, and $P_{\mathrm{De}} \in \{1.5,1.75,2.0\}$, where $P_{\mathrm{De}}=1.75$ serves as an intermediate setting. For PDPTW, which has no ride-time limits, we use the same instances but set the ride-time limit to 100 times the direct travel time. Unlike in DARP, where $P_{\mathrm{De}}$ affects both time windows and ride-time limits, in PDPTW it affects only the time windows. We focus on tight temporal settings and high request intensity, as the current $P_{\mathrm{De}}$ values already capture the relative advantages of the procedures under different levels of tightness and prior work shows that TSFrag+C performs well under low intensity. We denote the variant using a time resolution of $x \in \{1,2,5\}$ minutes by TSFrag+C$(x\,\text{min})$, where larger $x$-values reduce the model size at the cost of temporal accuracy.

All our algorithms are implemented using the Python programming language. All computational experiments are run on a system with an Intel Core i7-7820HQ processor with 2.90 GHz CPU speed and 32 GB of RAM under a Windows 10 64-bit OS\@. All linear formulations are solved with the commercial solver Gurobi 9.0.3 with a single thread; all other Gurobi parameters are set to their default values.

All procedures are executed under a 30-minute time limit, and all performance indicators are obtained within this limit. $N_{\text{Opt}}$ denotes the number of instances solved to optimality for EBF, TSEF+DDD, and TSFrag+DDD\@. For EBF and TSFrag+C, we additionally report the relative gap between the best objective value and the best-known lower bound, denoted as ``Gap". For TSEF+DDD and TSFrag+DDD, $Iter$ indicates the iteration count required for DDD convergence.
``Time'' is the total runtime (s) of each procedure. The values $|V_E|$, $|A_E|$, and $|F|$ represent the number of event nodes, event arcs, and (physical) fragments.

\subsection{Computational performance of different procedures for DARP-SV}\label{ch42}

\begin{table}[h]
	\centering
	\footnotesize
	\begin{threeparttable}
		\setlength{\tabcolsep}{0.5 pt}
		\caption{Computational results of different procedures for DARP-SV}
		\label{resultsdarpsv}
		\begin{tabular}{cccccccccccccccc}
			\toprule
			&                      &                     &          & \multicolumn{5}{c}{EBF}                                                     & \multicolumn{3}{c}{TSEF+DDD}                               & \multicolumn{4}{c}{TSFrag+DDD}             \\ \midrule
			Dataset                                                                                 & $R_L$                & $P_{TW}$            & $P_{De}$ & $|V_E|$ & $|A_E|$ & Time   & $N_{\text{Opt}}$ & \multicolumn{1}{c|}{Gap}    & Time   & $N_{\text{Opt}}$ & \multicolumn{1}{c|}{$Iter$} & $|F|$  & Time  & $N_{\text{Opt}}$ & $Iter$ \\ \midrule
			Set 1,   $|V|  \times 3$                                                                & 1/3                  &                     &          & 316.0   & 1822.8  & 1.2    & 48               & \multicolumn{1}{c|}{0.00\%} & 4.9    & 48               & \multicolumn{1}{c|}{5.6}    & 620.5  & 3.1   & 47               & 2.7    \\ \midrule
			\multirow{6}{*}{\begin{tabular}[c]{@{}c@{}}Set 2,\\      $|V|  \times 4$\end{tabular}} & \multirow{4}{*}{1/3} & \multirow{2}{*}{15} & 1.5      & 168.5   & 1006.6  & 60.6   & 13               & \multicolumn{1}{c|}{0.00\%} & 69.1   & 13               & \multicolumn{1}{c|}{18.9}   & 69.6   & 26.1  & 13               & 11.5   \\
			&                      &                     & 2.0      & 1073.5  & 2637.8  & 177.0  & 12               & \multicolumn{1}{c|}{0.06\%} & 616.5  & 10               & \multicolumn{1}{c|}{22.2}   & 425.0  & 23.7  & 13               & 11.9   \\ \cmidrule{3-16} 
			&                      & \multirow{2}{*}{30} & 1.5      & 248.6   & 1684.3  & 750.0  & 8                & \multicolumn{1}{c|}{0.58\%} & 559.0  & 11               & \multicolumn{1}{c|}{21.5}   & 87.5   & 390.3 & 13               & 14.8   \\
			&                      &                     & 2.0      & 2308.1  & 8917.1  & 1105.8 & 6                & \multicolumn{1}{c|}{4.66\%} & 1306.7 & 4                & \multicolumn{1}{c|}{15.3}   & 1292.8 & 788.7 & 8                & 14.6   \\ \cmidrule{2-16} 
			& \multirow{2}{*}{1/6} & \multirow{2}{*}{15} & 1.5      & 221.5   & 1069.3  & 2.3    & 13               & \multicolumn{1}{c|}{0.00\%} & 21.8   & 13               & \multicolumn{1}{c|}{15.5}   & 83.8   & 9.6   & 13               & 8.5    \\
			&                      &                     & 2.0      & 1855.2  & 4096.6  & 326.5  & 12               & \multicolumn{1}{c|}{0.14\%} & 864.1  & 7                & \multicolumn{1}{c|}{20.0}   & 689.7  & 14.5  & 13               & 8.4    \\ \midrule
			\multirow{2}{*}{\begin{tabular}[c]{@{}c@{}}Set 2,\\      $|V|  \times 5$\end{tabular}} & \multirow{2}{*}{1/3} & \multirow{2}{*}{15} & 1.5      & 168.5   & 1006.6  & 1.9    & 13               & \multicolumn{1}{c|}{0.00\%} & 16.4   & 13               & \multicolumn{1}{c|}{15.8}   & 69.6   & 9.5   & 13               & 8.8    \\
			&                      &                     & 2.0      & 1073.5  & 2637.8  & 179.0  & 12               & \multicolumn{1}{c|}{0.06\%} & 632.1  & 10               & \multicolumn{1}{c|}{22.4}   & 425.0  & 15.6  & 13               & 10.8   \\ \bottomrule
		\end{tabular}
		\begin{tablenotes}
			\item[1] All values are averages, except the $N_{\text{Opt}}$ column, which reports the number of instances solved to optimality. 
			\item[2] ``$|V|  \times 3$'' means the instance uses three times the number of vehicles of \citeauthor{doi:10.1287/opre.1060.0283}; similarly for ``$|V|  \times 4$'' and ``$|V|  \times 5$''.
		\end{tablenotes}
	\end{threeparttable}
\end{table}

Table~\ref{resultsdarpsv} reports the average results of the three procedures over the two benchmark sets, with detailed outcomes in Appendix~\ref{CompResults:darpsv}. ABF is excluded due to its poor performance (see Appendix~\ref{CompResults:darpsv}). Two key findings emerge for DARP-SV.

First, regarding the formulations, TSFrag+DDD performs best under high request density (the second dataset), whereas EBF performs best under low request density (the first dataset). TSFrag+DDD solves more instances to optimality and is faster in the second dataset (see the columns of Time and $N_{Opt}$), while EBF shows these advantages in the first. EBF's average runtime grows from 1.2 seconds in the first dataset to several hundred seconds in the second set; this might be because intense requests bring a heavy burden to its big-$M$ time-increment constraints. Importantly, rather than tight time windows, we find that high request density  is the main driver of TSFrag's advantage: in the last two rows, EBF is better when $P_{\text{De}}=1.5$, whereas TSFrag is better when $P_{\text{De}}=2.0$. Overall, TSFrag performs well under both high request density and tight temporal constraints; wider time temporal constraints increase fragment counts and prolong computation time, but they degrade EBF's performance even more severely.


Second, the fragment-based formulation aligns better with DDD than the event-based one: TSFrag+DDD outperforms TSEF+DDD in both solution time and iteration count (see the columns of $N_{\text{Opt}}$, Time, and $Iter$). This advantage arises because TSFrag naturally reduces rounding errors: each fragment might include multiple arcs, thereby mitigating discretization errors introduced during rounding. Additional evidence appears in Table~\ref{detailsresultsdarpsv13} of Appendix~\ref{CompResults:darp}, where TSFrag~(1~min) consistently attains better objective values than TSEF~(1~min); these runs correspond to the initial one-minute-resolution iterations of TSFrag+DDD and TSEF+DDD\@. This confirms that TSFrag+DDD provides tighter lower bounds and explains why it requires fewer iterations.

For completeness, we note that direct comparisons between TSFrag and TSFrag+DDD (or between TSEF and TSEF+DDD) are not presented, as the former produce discrete-time solutions while the latter yield continuous-time solutions, making their objective values not directly comparable. Nevertheless, detailed information on all four procedures is provided in Tables~\ref{detailsresultsdarpsv1} and \ref{detailsresultsdarpsv13} of Appendix~\ref{CompResults:darpsv} (corresponding to the first and second dataset, respectively), where TSFrag and TSEF are implemented with a 1-minute discretization.

\subsection{Computational performance of different procedures for DARP and PDPTW}\label{ch43}

\begin{table}[htbp]
	\centering
	\scriptsize
	\begin{threeparttable}
	\setlength{\tabcolsep}{1.0 pt}
	\caption{Computational results of different procedures for DARP and PDPTW}
	\label{resultsdarp}
	\begin{tabular}{@{}ccccccccccccccccc@{}}
		\toprule
		&          & \multicolumn{4}{c}{EBF}                                 & \multicolumn{3}{c}{TSFrag+DDD}               & \multicolumn{4}{c}{TSFrag+C (1 min)}               & \multicolumn{2}{c}{2 min}           & \multicolumn{2}{c}{5 min} \\ \midrule
		& $P_{De}$ & $|V_E|$ & $|A_E|$ & Time  & \multicolumn{1}{c|}{Gap}    & $|F|$  & Time  & \multicolumn{1}{c|}{$Iter$} & Time  & Gap    & NC  & \multicolumn{1}{c|}{NC/F}   & Time  & \multicolumn{1}{c|}{Gap}    & Time        & Gap         \\ \midrule
		\multirow{3}{*}{\begin{tabular}[c]{@{}c@{}}DARP\\      (24)\end{tabular}}  & 1.5      & 359.5   & 1299.0  & 8.1   & \multicolumn{1}{c|}{0.00\%} & 112.9  & 7.9   & \multicolumn{1}{c|}{9.1}    & 65.3  & 0.00\% & 2.9 & \multicolumn{1}{c|}{2.61\%} & 84.2  & \multicolumn{1}{c|}{0.00\%} & 158.0       & 0.03\%      \\
		& 1.75     & 1609.4  & 3251.4  & 259.1 & \multicolumn{1}{c|}{0.29\%} & 387.4  & 19.5  & \multicolumn{1}{c|}{10.3}   & 56.1  & 0.00\% & 3.0 & \multicolumn{1}{c|}{0.76\%} & 60.5  & \multicolumn{1}{c|}{0.00\%} & 235.7       & 0.04\%      \\
		& 2.0      & 4385.7  & 9990.6  & 692.0 & \multicolumn{1}{c|}{2.34\%} & 1607.3 & 96.9  & \multicolumn{1}{c|}{10.3}   & 156.5 & 0.03\% & 4.1 & \multicolumn{1}{c|}{0.25\%} & 147.2 & \multicolumn{1}{c|}{0.06\%} & 138.7       & 0.12\%      \\ \midrule
		\multirow{3}{*}{\begin{tabular}[c]{@{}c@{}}DARP\\      (16)\end{tabular}}  & 1.5      & 169.7   & 625.9   & 1.0   & \multicolumn{1}{c|}{0.00\%} & 65.1   & 1.5   & \multicolumn{1}{c|}{6.7}    & 2.9   & 0.00\% & 1.6 & \multicolumn{1}{c|}{2.50\%} & 2.5   & \multicolumn{1}{c|}{0.00\%} & 2.6         & 0.00\%      \\
		& 1.75     & 537.9   & 1139.6  & 13.3  & \multicolumn{1}{c|}{0.00\%} & 165.3  & 4.1   & \multicolumn{1}{c|}{7.5}    & 5.7   & 0.00\% & 1.4 & \multicolumn{1}{c|}{0.87\%} & 5.6   & \multicolumn{1}{c|}{0.00\%} & 11.9        & 0.00\%      \\
		& 2.0      & 1352.7  & 2738.7  & 210.1 & \multicolumn{1}{c|}{0.22\%} & 465.1  & 8.1   & \multicolumn{1}{c|}{7.3}    & 7.3   & 0.00\% & 1.7 & \multicolumn{1}{c|}{0.36\%} & 5.9   & \multicolumn{1}{c|}{0.00\%} & 6.2         & 0.00\%      \\ \midrule
		\multirow{3}{*}{\begin{tabular}[c]{@{}c@{}}PDPTW\\      (16)\end{tabular}} & 1.5      & 3365.2  & 5469.3  & 390.8 & \multicolumn{1}{c|}{0.52\%} & 1633.5 & 43.6  & \multicolumn{1}{c|}{9.9}    & 24.1  & 0.00\% & 2.2 & \multicolumn{1}{c|}{0.13\%} & 21.4  & \multicolumn{1}{c|}{0.00\%} & 28.4        & 0.00\%      \\
		& 1.75     & 5045.2  & 9489.4  & 472.8 & \multicolumn{1}{c|}{0.66\%} & 3568.8 & 56.5  & \multicolumn{1}{c|}{7.7}    & 39.9  & 0.00\% & 1.5 & \multicolumn{1}{c|}{0.04\%} & 36.1  & \multicolumn{1}{c|}{0.00\%} & 38.7        & 0.00\%      \\
		& 2.0      & 6978.1  & 15108.6 & 710.1 & \multicolumn{1}{c|}{1.00\%} & 7155.4 & 100.2 & \multicolumn{1}{c|}{8.4}    & 76.6  & 0.00\% & 0.8 & \multicolumn{1}{c|}{0.01\%} & 71.5  & \multicolumn{1}{c|}{0.00\%} & 72.3        & 0.00\%      \\ \bottomrule
	\end{tabular}
	\begin{tablenotes}
		\item[1] All values are averages. All instances solved by TSFrag+DDD reach optimality. ``2~min'' and ``5~min'' refer to TSFrag+C with 2- and 5-minute resolutions. 
		\item[2] ``DARP (24)'' and ``DARP (16)'' report averages over all 24 DARP instances and the first 16, respectively; ``PDPTW (16)'' reports averages over the first 16 PDPTW instances (larger cases are omitted as they as they are difficult to solve to optimality).
	\end{tablenotes}
\end{threeparttable}
\end{table}

Table~\ref{resultsdarp} reports the average computational results of EBF, TSFrag+DDD, and TSFrag+C for DARP and PDPTW, with detailed results in Appendix~\ref{CompResults}. Here, we also introduce the indicators ``NC'' and ``NC/F'' to record the number of callbacks and their ratio to the fragment count. The key findings are summarized below. 

First, we find that for DARP, TSFrag+DDD generally outperforms EBF under high request intensity, and the Gap grows as temporal constraints widen. For DARP~(24), EBF performs similarly to TSFrag+DDD at $P_{De}=1.5$ but becomes markedly slower at $P_{De}=1.75$ and $P_{De}=2.0$. Both formulations face a larger event/fragment count and longer runtime under wider time windows, but EBF is affected more strongly.

Second, for DARP, we find that DDD can effectively replace callbacks. From DARP~(24), TSFrag+DDD consistently outperforms TSFrag+C with 1-, 2-, and 5-minute resolutions in solution time, with the advantage particularly pronounced under tight temporal constraints due to increased callback pressure. This pattern aligns with \citet{sippel2024}, who observed similar behavior for TSFrag+DDD on PDPTW\@. For TSFrag+C, the 1-minute resolution performs best at $P_{De}=1.5$, whereas the 5-minute resolution performs best at $P_{De}=2.0$. This indicates that small time resolutions suit tight temporal settings and larger ones suit wider settings, and therefore additional resolutions are not tested.

Third, we find that DARP benefits more from DDD than PDPTW\@. Given that PDPTW has far more events, event arcs, and fragments, only the first 16 instances are used for comparison, as the remaining larger instances are typically hard to solve optimally and thus the total NC required to reach optimality cannot be reported.
All three procedures involve a larger number of variables and require longer runtimes when moving from DARP to PDPTW.  The procedure TSFrag+DDD is among the fastest for DARP~(16) and DARP~(24), yet for PDPTW~(16) it is slower than TSFrag+C with 1-, 2-, and 5-minute resolutions. By counting NC and NC/F, we observe that NC/F drops sharply in PDPTW compared with DARP\@. Since DDD is used to replace callbacks, its advantage diminishes when callbacks become rare, thereby explaining why DARP benefits more than PDPTW\@. \citet{sippel2024} report strong  performance for TSFrag+DDD on PDPTW, but because they employ additional techniques and use different instance sets, the experimental settings are not directly comparable; our intention is not to contradict their findings but to highlight that the inherent ride-time constraint in DARP leads to more frequent callbacks, thereby making DDD more impactful.

\section{Conclusions}\label{ch4:conclusion}

In this study, we have introduced the Dial-a-Ride Problem with Synchronized Visits, motivated by drone delivery and modular vehicle systems where limited vehicle capacity may require multiple units to serve a large customer simultaneously. We have  developed event-based, time-space event-based, and time-space fragment-based formulations by extending the event-based method of \citet{Gaul2024} and the fragment-based framework of \citet{Alyasiry2019}. The main extension lies in enabling synchronized visits. We have also adapted  the DDD algorithm to this problem within the time-space formulations. Computational results show that for DARP-SV, the event-based formulation performs best under normal time-window and ride-time settings with low request intensity, whereas the time-space fragment-based formulation with DDD excels under tight temporal settings and high request intensity. The results also show that DDD aligns more naturally with fragment-based than with event-based formulations. Additionally, for the classical DARP, DDD can effectively replace callbacks 
for the time-space fragment-based formulation under high request intensity and provides greater benefits than for PDPTW.


\bibliographystyle{informs2014_doi}
\bibliography{library2}

\begingroup \parindent 0pt \parskip 0.0ex \def\enotesize{\normalsize} \endgroup
\clearpage

\appendix

\small

\section{Arc-based formulation for DARP-SV}\label{ABFappendix}

This section details the arc-based formulation ABF for DARP-SV\@. The formulation below is largely the same as the DARP formulation from \citet{doi:10.1287/opre.1060.0283} but it incorporates large customers and synchronisation.
All decision variables are summarized in Table~\ref{table:ch4notation3.1}. We also define the new parameter $vq_i$ that denotes the load variation of a single vehicle when serving location $i$. For a small customer~$i\in P_s$ we set $vq_i = q_i$ at its pickup location and $vq_{i+n} = -q_i$ at its delivery location, while for a large customer~$i\in P_l$ we have $vq_i = Q$ and $vq_{i+n} = -Q$.

\begin{table}[h]
\footnotesize
\centering
\caption{Variables of arc-based formulation}
\begin{tabular}{@{}ll@{}}
	\toprule
	Decision variables 						 &Definition                   \\\midrule
	$f_{ijv}$                                  & = 1, if the location arc  $(i,j)\in A$ is traversed by vehicle $v\in V$; = 0, otherwise\\
	$tt_{i}$                                  & the departure time at a large customer's location $i\in P_l \cup D_l$\\
	$t_{iv}$                                  & the departure time of vehicle $v\in V$ at location $i\in N$\\
	$Q_{iv}$                                  & the load of vehicle $v$ after visiting $i\in N$\\
	\bottomrule
\end{tabular}
\label{table:ch4notation3.1}
\end{table}

\allowdisplaybreaks
\begin{align}
&& \min\sum_{v\in V} \sum_{(i,j)\in A} C_{ij} f_{ijv}\label{ch4:DARPLobj1}\\
s.t.
&&\sum_{v\in V} \sum_{j\in N}f_{ijv}                  &= \lceil q_i/Q \rceil    &&i\in N\label{ch4:cover2}\\
&&\sum_{j \in N}f_{ijv}  &=\sum_{j\in N}f_{(i+n)jv}      &&i\in P, v\in V  \label{ch4:pair3}\\
&&\sum_{j\in N}f_{0jv}  &=1      &&v\in V  \label{ch4:flow4}\\
&&\sum_{j\in N}f_{jiv}  &=\sum_{j\in N}f_{ijv}     &&i\in P\cup D, v\in V  \label{ch4:flow5}\\
&&\sum_{i\in N}f_{i(2n+1)v}  &=1      &&v\in V  \label{ch4:flow6}\\
&&t_{iv}  &= tt_{i}     &&i\in P_l \cup D_l, v\in V  \label{relatesupport}\\
&&t_{iv}+ T_{ij}  &\le t_{jv} +M_{ij}(1-f_{ijv})     &&(i,j)\in A, v\in V  \label{ch4:time7}\\
&&e_i \le t_{iv} &\le l_i     &&i\in N,  v\in V \label{ch4:time8}\\%
&&Q_{iv}+vq_{j}  &\le Q_{jv} + W_{ij}(1-f_{ijv})     &&(i,j)\in A, v\in V \label{ch4:capacity9}\\ 
&& Q_{iv} &= Q && i\in P_l,\; v\in V \label{ch4:largecap1}\\
&& Q_{(i+n)v} &= 0 && i\in P_l,\; v\in V \label{ch4:largecap2}\\
&&\max(0,vq_i) \le Q_{iv} &\le \min(Q,Q+vq_i)      &&i\in N,  v\in V  \label{ch4:capacity10}\\
&&t_{(n+i)v}- t_{iv} &\le R_i      &&i\in P, v\in V   \label{ch4:maxride11}\\
&&f_{ijv} 				&\in\{0, 1\}	&&(i,j)\in A, v\in V \label{ch4:domain}\\
&&t_{iv}				&\ge 0	&&i\in N, v\in V \label{ch4:domain1}\\
&&tt_{i}				&\ge 0	&&i\in P_l\cup D_l \label{ch4:domain2}\\
&&Q_{iv}				&\in\mathbb{N}	&&i\in N, v\in V \label{ch4:domain3}
\end{align}	

The objective function (\ref{ch4:DARPLobj1}) aims to minimize the total cost of traversing location arcs. Constraints~(\ref{ch4:cover2}) ensure that the pickup location of each small customer is visited by a single vehicle, and the pickup location of each large customer is visited by $\lceil q_i/Q \rceil$ vehicles. Constraints~(\ref{ch4:pair3}) state that paired locations are traversed by the same vehicle. Constraints~(\ref{ch4:flow4}) to (\ref{ch4:flow6}) specify the network flow requirements. Constraints~(\ref{relatesupport}) ensure that the departure times of multiple vehicles visiting the same large customer are synchronized at the customer's pickup and delivery location. Specifically, we ensure that all vehicles visiting the large customer's location~$i \in P_l \cup D_l$ depart simultaneously at time $tt_i$, thereby synchronizing departures from location $i$. Constraints~(\ref{ch4:time7}) to (\ref{ch4:capacity10}) ensure time and load consistency, where $M_{ij} = \max(0, l_i + T_{ij} - e_j)$ and $W_{ij} = \min(Q, Q + vq_i)$. Constraints~(\ref{ch4:largecap1}) and~(\ref{ch4:largecap2}) enforce that, after serving the pickup and delivery location of a large customer~$i$, the vehicle load equals $Q$ and $0$, respectively. Constraint~(\ref{ch4:capacity10}) bounds the feasible load range at each location~$i$. Constraints~(\ref{ch4:maxride11}) set the maximum ride time for customers. Constraints~(\ref{ch4:domain}) to (\ref{ch4:domain3})  specify the domain of the decision variables $f$, $t$, $tt$, and $Q$.

\section{Details of DDD}\label{DDDapply}

This section provides additional details on DDD, including the construction of partial time-space networks (Section~\ref{DDDstep1}) and the selection model (Section~\ref{DDDselctionmodel}).

\subsection{The construction of a partial time-space network}\label{DDDstep1}

\textbf{Step~1} of the DDD procedure constructs a partial time-space network $G(\mathcal{T}^k)$ using iteration~$k$'s time index set $\mathcal{T}^k$ and formulates the relaxed problem~$\mathcal{P}_{\mathcal{T}^k}$. 
In the initial iteration ($k = 0$), $\mathcal{T}^k_p$ includes only the earliest departure and latest arrival times at node $p$. 
For $k > 1$, $\mathcal{T}^k$ is carried over from iteration $k - 1$. 
Each node~$p$ has its own time set $\mathcal{T}^k_p$, and $\mathcal{T}^k$ is the union of all $\mathcal{T}^k_p$. 
Because $\mathcal{T}^k$ does not contain all necessary time indices, constructing $G(\mathcal{T}^k)$ requires adjustments compared to Section~\ref{ch4:fragmentformulate}. When generating a time-space fragment (or node arc) with departure time $t$, the earliest arrival time $t'$ at the end node (calculated according to the earliest schedule along the corresponding route path) is rounded down to the nearest time index in $\mathcal{T}^k_{p'}$, denoted $\mathrm{round}(t')$. The shortened arc time length ($\mathrm{round}(t') - t$) is therefore strictly smaller than the actual arc time length ($t' - t$), and may even become negative if the discretization is too coarse. A comparison between the full and partial time-space networks is illustrated in Fig.~\ref{TSRFBNBasic} and Fig.~\ref{TSRFBNround}, respectively, with the latter also illustrating the rounding issue. Consider the arc $(p7, d7)$ as an example. In Fig.~\ref{TSRFBNBasic}, the arrival time at $d7$ belongs to the full time set, while in Fig.~\ref{TSRFBNround}, this arrival time does not belong to the current coarse time set. In this case, the arc $(p7, d7)$ departs from purple $p7$ at $t = 1$, has a travel time of $T_{p7d7} = 1.3$, and arrives at blue $d7$ at $t' = 2.3$. After rounding, $\mathrm{round}(t') = 2$ (purple $d7$), and the shortened arc time length becomes $1$ ($=2-1$).

\subsection{The selection model}\label{DDDselctionmodel}

\textbf{Step~3} of the DDD method evaluates whether the paths derived from $S(\mathcal{P}_{\mathcal{T}^k})$ are feasible in $G(\mathcal{T})$  using a selection model. 
Unlike \citet{Boland2017}, our selection model is tailored for DARP-SV and incorporates additional constraints to account for fragment structures.

Before introducing the selection model, we define several parameters derived from $S(\mathcal{P}_{\mathcal{T}^k})$. 
We denote the sets of used vehicles, traversed fragments, and traversed node arcs as $\bar{V}$, $\bar{F}$, and $\bar{A}_N$, respectively. 
If a location arc $(i,j)$ belongs to a fragment or node arc~$f$, we denote this relationship by $(i,j) \in f$.  
Let $P_v = \{i_1 = 0, i_2, \dots, i_p, 2n+1\}$ represent the ordered sequence of locations visited by vehicle $v$ in $S(\mathcal{P}_{\mathcal{T}^k})$, and let $P'_v$ denote the corresponding set of location arcs $(i_k,i_{k+1})$, where $i_k,i_{k+1}\in P_v$. 
Finally, $\bar{T}_f$ denotes the shortened time length of fragment $f \in \bar{F}$, obtained from the corresponding time-space fragment containing~$f$. 

We now introduce a key parameter, the shortened time length $\bar{T}_{ij}$ for location arc $(i,j)$, which differs from $\bar{T}_f$ for fragment $f$. In the time-space arc-based network \citep{Boland2017}, $\bar{T}_{ij}$ is directly obtained from the time length of each time-space arc. In TSFrag, however, a fragment may contain multiple location arcs, and only the fragment’s shortened length $\bar{T}_{f}$ is available from $G(\mathcal{T}^{k})$. Hence, $\bar{T}_{ij}$ must be derived from $\bar{T}_{f}$. Let $t'$ be the actual fragment end time and $\mathrm{round}(t')$ its rounded value; the value $(t'-\mathrm{round}(t'))$ is the timing discrepancy from discretization. For each location arc~$(i,j)$, we compute $\bar{T}_{ij} = T_{ij} - \bigl(t' - \mathrm{round}(t')\bigr)$ where the full discrepancy is assigned to each location arc. This aims to ensure time-window feasibility, since each location has its own time window and the required share for each location arc cannot be determined in advance. Moreover, to ensure that each fragment contains at most one discrepancy, we enforce constraints~\eqref{identifyfragminimunlength}. For node arcs in TSFrag (or event arcs in TSEF), the corresponding location arc has the same shortened time length. 

The decision variables of the selection model are summarized in Table~\ref{tab:selection_decision_variables}. We also let $\tau_{\mathrm{start}(f)}$ and $\tau_{\mathrm{end}(f)}$ denote the start and end nodes of fragment $f$, respectively.

\begin{table}[htbp]
\centering
\footnotesize
\caption{Decision variables in the selection model}
\label{tab:selection_decision_variables}
\begin{tabular}{ll}
	\hline
	Variables &Description \\ 
	\hline
	$\tau_i$ & Departure time at location $i \in N$; no vehicle index is required because no more than \\& one vehicle traverses the same location at different times. \\[0.3em]
	$\theta_{ij}$ & $\theta_{ij} \ge 0$, effective travel time used on location arc $(i,j) \in P'_v$, $v \in \bar{V}$. \\[0.3em]
	$\delta_{ij}$ & $=1$ if location arc $(i,j)$ is allowed to use a travel time shorter than its actual arc \\& time length; $=0$ otherwise. \\
	\hline
\end{tabular}
\end{table}

The selection model based on $S(\mathcal{P}_{\mathcal{T}^k})$ is as follows.  

{\small
\setlength{\abovedisplayskip}{5pt}
\setlength{\belowdisplayskip}{5pt}
\setlength{\jot}{2pt}
\allowdisplaybreaks
\begin{align}
	Z = \min\; & \sum_{v\in \bar{V}} \sum_{(i,j)\in P'_v} \delta_{ij}
	\label{identifyobj}\\[0.3em]
	\text{s.t.}\quad
	& \theta_{ij} \ge T_{ij}(1-\delta_{ij}),
	&& \forall (i,j)\in P'_v,\ v\in\bar{V} \label{identifyrelate}\\
	& \tau_i + \theta_{ij} \le \tau_j,
	&& \forall (i,j)\in P'_v,\ v\in\bar{V} \label{identifyincrement}\\
	& \tau_{\mathrm{end}(f)} - \tau_{\mathrm{start}(f)} 
	\ge \bar{T}_f,
	&& \forall f\in \bar{F} \label{identifyfragminimunlength}\\
	& \theta_{ij} \ge \bar{T}_{ij},
	&& \forall (i,j)\in P'_v,\ v\in\bar{V} \label{identifyminimunlength}\\
	& e_i \le \tau_i \le l_i,
	&& \forall i\in N \label{identifytw}\\
	& \delta_{ij} \in \{0,1\},\; \theta_{ij}\ge 0,
	&& \forall (i,j)\in P'_v,\ v\in\bar{V} \label{identifydomain}
\end{align}
}

The objective~(\ref{identifyobj}) is to minimize the number of arcs that require an arc time length shorter than the actual arc time length. Constraints~(\ref{identifyrelate}) specify that if a location arc has a travel time $\theta$ shorter than the actual arc time length, its corresponding variable $\delta$ is set to 1. Constraints~(\ref{identifyincrement}) ensure that the time increment is based on the travel time $\theta$. Constraints~(\ref{identifyfragminimunlength}) specify that for a fragment containing one or more location arcs, the difference between the departure time at the end node and the departure time at the start node must be greater than or equal to the fragment's shortened time length. This requirement arises because the sum of the shortened arc time lengths of all location arcs is less than the shortened fragment time length $\bar{T}_f$, and $\bar{T}_f$ more accurately reflects the time length parameter in $G(\mathcal{T}^{k})$. Additionally, constraints~(\ref{identifyminimunlength}) ensure that the travel time $\theta$ is not less than the shortened arc time length. Constraints~(\ref{identifytw}) define the earliest and latest times for each location, while constraints~(\ref{identifydomain}) set the domain of $\delta$.



\section{Computational details and results for DARP-SV and DARP}\label{CompResults}

\subsection{Details of instance preprocessing and construction}\label{CompResults:preprocessing}

We tighten the original time windows to eliminate infeasible temporal intervals induced by depot reachability and ride-time consistency, following \citet{Rist2021}. For each customer $i\in P$, a single forward-backward propagation applies the following three steps. (1) Tightening pickup windows: the earliest feasible pickup time becomes $e_i\leftarrow\max\{e_i,\;e_{i+n}-R_i,\;e_0+t_{0i}\}$. (2) Tightening delivery windows: the latest feasible delivery time becomes $l_{i+n}\leftarrow\min\{l_{i+n},\;l_i+R_i,\;l_{2n+1}-t_{(i+n)(2n+1)}\}$. (3) Enforcing pickup-delivery consistency; the updated delivery window refines both sides: $l_i\leftarrow\min\{l_i,\;l_{i+n}-t_{i(i+n)}\}$ and $e_{i+n}\leftarrow\max\{e_{i+n},\;e_i+t_{i(i+n)}\}$.

After this tightening, all earliest pickup times are mapped into a single hour by $e_i\leftarrow e_i\bmod 60$ to induce high request density, and the latest pickup time is set to $l_i=e_i+P_{TW}$ for $P_{TW}\in\{15,30\}$. Delivery windows are then reconstructed using the direct travel time $t_{i,i+n}$ and the maximum ride-time factor $P_{\mathrm{De}}\in\{1.5,2.0\}$, yielding $e_{i+n}=e_i+t_{i,i+n}$ and $l_{i+n}=e_i+P_{\mathrm{De}}\,t_{i,i+n}$, which ensures consistency with both travel-time feasibility and the prescribed ride-time limit. Finally, the entire time horizon is shifted forward by 30 minutes and the destination depot's latest allowable time is set to a sufficiently large value to avoid infeasible time schedules.

\subsection{Computational results for DARP-SV}\label{CompResults:darpsv}

This section provides all the details of the computational results for DARP-SV.


For DARP-SV, we present the computational results of EBF, TSEF+DDD, TSFrag+DDD, TSEF (1~min), and TSFrag (1~min) on the first dataset in Table~\ref{detailsresultsdarpsv1}. Here, TSEF (1~min) and TSFrag (1~min) refer to  TSEF and TSFrag run with a one-minute resolution for a single round, providing an approximate solution and a corresponding lower bound. We also present the results of EBF, TSEF+DDD, and TSFrag+DDD for the second dataset with varying parameters reported in Tables~\ref{detailsresultsdarpsv13}-\ref{detailsresultsdarpsv16}. Computational results on the 
instances of the first benchmark solved to optimality by ABF are summarized in Table~\ref{detailsresultsdarpsabf}. Given its substantially inferior performance relative to the other formulations, we omit the detailed results of ABF for the remaining instances.

Apart from the indicators described in Section~\ref{ch41}, several additional indicators are reported under the 30-minute time limit. For EBF, ``OBJ'' and ``LB'' denote the best-found objective value and the best-known lower bound within this limit. 
For TSEF (1~min) and TSFrag (1~min), ``OBJ-A'' and ``LB-A'' denote the best-found objective value and the best-known lower bound for the approximate problem with a 1-minute time accuracy. For TSFrag+DDD, TSFrag (1~min), and TSFrag ($x$~min)+C, the indicator ``$|F|$'' is identical. For TSEF+DDD and TSFrag+DDD, we use ``OBJ/LB'' to denote the best-known lower-bound value of the original problem. When the corresponding ``Time'' value is less than 1800~seconds,  ``OBJ/LB'' also represents the optimal objective value. In addition, ``Avg'' represents the average value of each column. However, in the ``OBJ'', ``LB'', and ``OBJ/LB'' columns, the value in the ``Avg'' row has a different meaning: it reports the total number of instances solved to optimality. ``NA'' in the following table indicates that no feasible solution was obtained within the time limit. The remaining definitions follow those used in the body of this manuscript.

\clearpage
\begin{table}[H]
\centering
\scriptsize
\renewcommand{\arraystretch}{0.4}  
\caption{Results for the instances in the first dataset solved to optimality  by ABF for DARP-SV}
\label{detailsresultsdarpsabf}
\begin{tabular}{@{}lcc@{}}
	\toprule
	Name  & \multicolumn{1}{l}{OBJ} & \multicolumn{1}{l}{Time} \\ \midrule
	a2-16 & 456.99                  & 1.9                      \\
	a2-20 & 461.28                  & 8.4                      \\
	a2-24 & 581.19                  & 14.6                     \\
	a3-18 & 460.08                  & 24.0                     \\
	a3-24 & 493.26                  & 560.8                    \\
	a3-30 & 735.69                  & 1184.0                   \\
	a4-16 & 426.83                  & 80.9                     \\
	b2-16 & 383.79                  & 12.3                     \\
	b2-20 & 446.03                  & 4.1                      \\
	b2-24 & 575.61                  & 23.2                     \\
	b3-18 & 413.00                  & 57.1                     \\
	b3-24 & 505.61                  & 114.3                    \\
	b3-30 & 661.33                  & 228.0                    \\
	b3-36 & 759.29                  & 1230.8                   \\
	b4-16 & 430.97                  & 31.4                     \\
	b4-24 & 523.95                  & 1218.2                   \\ \bottomrule
\end{tabular}
\end{table}

\begin{table}[H]
\centering
\scriptsize
\setlength{\tabcolsep}{0.5pt}
\renewcommand{\arraystretch}{0.4}  
\caption{Computational results for the first dataset for DARP-SV}
\label{detailsresultsdarpsv1}
\begin{tabular}{@{}cccccccccccccccc@{}}
	\toprule
	& \multicolumn{4}{c}{EBF}         & \multicolumn{3}{c}{TSEF+DDD} & \multicolumn{4}{c}{TSFrag+DDD} & \multicolumn{2}{c}{TSEF (1 min)} & \multicolumn{2}{c}{TSFrag (1 min)} \\ \midrule
	Name  & E     & EA     & OBJ     & Time & Iter    & OBJ       & Time   & F      & Iter & OBJ     & Time & OBJ              & Time          & OBJ               & Time           \\
	a2-16 & 46    & 182    & 456.99  & 0.1  & 1       & 456.99    & 0.1    & 24     & 1    & 456.99  & 0.1  & 456.99           & 0.3           & 456.99            & 0.4            \\
	a2-20 & 65    & 282    & 461.28  & 0.1  & 1       & 461.28    & 0.2    & 31     & 1    & 461.28  & 0.2  & 461.28           & 0.3           & 461.28            & 0.4            \\
	a2-24 & 69    & 373    & 581.19  & 0.1  & 6       & 581.19    & 0.7    & 36     & 4    & 581.19  & 0.1  & 581.19           & 3.8           & 581.19            & 1.1            \\
	a3-18 & 68    & 256    & 460.08  & 0.1  & 1       & 460.08    & 0.1    & 39     & 1    & 460.08  & 0.1  & 460.08           & 0.5           & 460.08            & 0.4            \\
	a3-24 & 107   & 432    & 493.26  & 0.2  & 5       & 493.26    & 0.9    & 59     & 2    & 493.26  & 0.1  & 493.26           & 1.2           & 493.26            & 0.5            \\
	a3-30 & 93    & 540    & 735.69  & 0.1  & 1       & 735.69    & 0.2    & 46     & 1    & 735.69  & 0.1  & 735.69           & 0.8           & 735.69            & 0.7            \\
\bottomrule
	\end{tabular}\end{table}

\begin{sidewaystable}
\centering
\tiny
\setlength{\tabcolsep}{0.5pt}
\caption{Computational results for the second dataset for DARP-SV ($|V|  \times 4$, $R_L=1/3$, $P_{TW}=15$)}
\label{detailsresultsdarpsv13}
\begin{tabular}{cccccccccccccccccccccc}
	\toprule
	& \multicolumn{6}{c}{EBF}                                 & \multicolumn{3}{c}{TSEF+DDD}  &       & \multicolumn{3}{c}{TSFrag+DDD} & \multicolumn{4}{c}{TSEF (1 min)}    & \multicolumn{4}{c}{TSFrag (1 min)} \\ \midrule
	& $|V_E|$ & $|A_E|$ & Time   & OBJ     & LB      & Gap    & Time   & OBJ/LB  & $Iter$ & $|F|$ & Time    & OBJ/LB    & $Iter$   & Time   & OBJ     & LB      & Gap    & Time  & OBJ     & LB      & Gap    \\\midrule
	$P_{De}=1.5$ &         &         &        &         &         &        &        &         &        &       &         &           &          &        &         &         &        &       &         &         &        \\\midrule
	a3-18        & 48      & 159     & 0.0    & 475.94  & 475.94  & 0.00\% & 0.4    & 475.94  & 9      & 23    & 0.2     & 475.94    & 4        & 0.3    & 473.13  & 473.13  & 0.00\% & 0.3   & 473.13  & 473.13  & 0.00\% \\
	a3-24        & 61      & 287     & 0.1    & 514.29  & 514.29  & 0.00\% & 0.6    & 515.66  & 8      & 30    & 0.2     & 514.29    & 3        & 0.6    & 513.82  & 513.82  & 0.00\% & 0.5   & 514.29  & 514.29  & 0.00\% \\
	a4-16        & 48      & 148     & 0.1    & 416.65  & 416.65  & 0.00\% & 0.5    & 416.65  & 9      & 22    & 0.1     & 416.65    & 3        & 0.2    & 416.65  & 416.65  & 0.00\% & 0.3   & 416.65  & 416.65  & 0.00\% \\
	a4-24        & 57      & 253     & 0.2    & 596.57  & 596.57  & 0.00\% & 0.7    & 589.53  & 2      & 27    & 0.2     & 596.57    & 2        & 0.6    & 596.62  & 596.57  & 0.00\% & 0.7   & 596.62  & 596.57  & 0.00\% \\
	a4-32        & 117     & 466     & 0.3    & 785.25  & 785.25  & 0.00\% & 3.7    & 785.25  & 16     & 49    & 1.2     & 785.25    & 7        & 3.6    & 785.25  & 785.25  & 0.00\% & 1.9   & 785.25  & 785.25  & 0.00\% \\
\bottomrule
	\end{tabular}\end{sidewaystable}

\begin{table}[H]
	\centering
	\scriptsize
	\setlength{\tabcolsep}{0.5 pt}
	\caption{Computational results for the second dataset for DARP-SV \\($|V| \times 4$, $R_L=1/3$, $P_{TW}=30$)}
	\label{detailsresultsdarpsv14}
	\begin{tabular}{@{}cccccccccccccc@{}}
		\toprule
		& \multicolumn{6}{c}{EBF}                                  & \multicolumn{3}{c}{TSEF+DDD}  &        & \multicolumn{3}{c}{TSFrag+DDD} \\ \midrule
		& $|V_E|$ & $|A_E|$ & Time   & OBJ     & LB      & Gap     & Time   & OBJ/LB  & $Iter$ & $|F|$  & Time     & OBJ/LB    & $Iter$  \\\midrule
		$P_{De}=1.5$ &         &         &        &         &         &         &        &         &        &        &          &           &         \\\midrule
		a3-18        & 53      & 256     & 0.5    & 433.08  & 433.08  & 0.00\%  & 14.5   & 433.08  & 24     & 25     & 3.9      & 433.08    & 10      \\
		a3-24        & 70      & 430     & 0.7    & 480.21  & 480.21  & 0.00\%  & 7.4    & 480.21  & 14     & 34     & 7.4      & 480.21    & 14      \\
		a4-16        & 48      & 193     & 0.1    & 386.03  & 386.03  & 0.00\%  & 0.7    & 386.03  & 11     & 22     & 0.6      & 386.03    & 8       \\
		a4-24        & 72      & 398     & 1.0    & 543.56  & 543.56  & 0.00\%  & 4.9    & 543.56  & 16     & 34     & 4.9      & 543.56    & 11      \\
		a4-32        & 141     & 756     & 11.8   & 710.63  & 710.62  & 0.00\%  & 59.1   & 710.63  & 24     & 58     & 25.1     & 710.63    & 15      \\
\bottomrule
	\end{tabular}
\end{table}

\begin{table}[H]
	\centering
	\scriptsize
	\setlength{\tabcolsep}{0.5pt}
	\caption{Computational results for the second dataset for DARP-SV \\($|V|  \times 4$, $R_L=1/6$, $P_{TW}=15$)}
	\label{detailsresultsdarpsv15}
	\begin{tabular}{@{}cccccccccccccc@{}}
		\toprule
		& \multicolumn{6}{c}{EBF}                                 & \multicolumn{3}{c}{TSEF+DDD}  &       & \multicolumn{3}{c}{TSFrag+DDD} \\ \midrule
		& $|V_E|$ & $|A_E|$ & Time   & OBJ     & LB      & Gap    & Time   & OBJ/LB  & $Iter$ & $|F|$ & Time    & OBJ/LB    & $Iter$   \\\midrule
		$P_{De}=1.5$ &         &         &        &         &         &        &        &         &        &       &         &           &          \\\midrule
		a3-18        & 53      & 166     & 0.1    & 411.24  & 411.24  & 0.00\% & 0.2    & 411.24  & 5      & 25    & 0.4     & 411.24    & 3        \\
		a3-24        & 66      & 297     & 0.1    & 466.00  & 466.00  & 0.00\% & 0.5    & 459.43  & 9      & 32    & 0.4     & 466.00    & 3        \\
		a4-16        & 48      & 148     & 0.1    & 356.30  & 356.30  & 0.00\% & 0.4    & 356.30  & 13     & 22    & 0.6     & 356.30    & 3        \\
		a4-24        & 64      & 262     & 0.1    & 518.52  & 518.52  & 0.00\% & 0.1    & 518.52  & 1      & 30    & 0.7     & 518.52    & 1        \\
		a4-32        & 142     & 504     & 0.3    & 650.04  & 650.04  & 0.00\% & 1.0    & 650.04  & 8      & 57    & 2.6     & 650.04    & 3        \\
\bottomrule
	\end{tabular}
\end{table}

\begin{table}[H]
\centering
\scriptsize
\setlength{\tabcolsep}{0.5pt}
\caption{Computational results for the second dataset for DARP-SV \\($|V|  \times 5$, $R_L=1/3$, $P_{TW}=15$)}
\label{detailsresultsdarpsv16}
\begin{tabular}{@{}cccccccccccccc@{}}
	\toprule
	& \multicolumn{6}{c}{EBF}                                 & \multicolumn{3}{c}{TSEF+DDD}  &       & \multicolumn{3}{c}{TSFrag+DDD} \\ \midrule
	& $|V_E|$ & $|A_E|$ & Time   & OBJ     & LB      & Gap    & Time   & OBJ/LB  & $Iter$ & $|F|$ & Time    & OBJ/LB    & $Iter$   \\\midrule
	$P_{De}=1.5$ &         &         &        &         &         &        &        &         &        &       &         &           &          \\\midrule
	a3-18        & 48      & 159     & 0.1    & 475.94  & 475.94  & 0.00\% & 0.5    & 475.94  & 9      & 23    & 0.4     & 475.94    & 4        \\
	a3-24        & 61      & 287     & 0.1    & 514.29  & 514.29  & 0.00\% & 0.5    & 515.66  & 8      & 30    & 0.7     & 514.29    & 3        \\
	a4-16        & 48      & 148     & 0.1    & 416.65  & 416.65  & 0.00\% & 0.4    & 416.65  & 9      & 22    & 0.5     & 416.65    & 3        \\
	a4-24        & 57      & 253     & 0.1    & 596.57  & 596.57  & 0.00\% & 0.1    & 589.53  & 2      & 27    & 0.7     & 596.57    & 2        \\
	a4-32        & 117     & 466     & 0.3    & 780.72  & 780.72  & 0.00\% & 1.5    & 780.72  & 13     & 49    & 2.5     & 780.72    & 5        \\
\bottomrule
\end{tabular}
\end{table}

\subsection{Computational results for DARP and PDPTW}\label{CompResults:darp}

\begin{sidewaystable}
	\centering
	\tiny
	\setlength{\tabcolsep}{0.2pt}
	\caption{Computational results for the second dataset for DARP with varying $P_{De}$ \\($|V|  \times 4$, $R_L=0$, $P_{TW}=15$, $P_{De}=1.5$)}
	\label{detailsresultsdarp}
	\begin{tabular}{@{}cccccccccccccccccccccccccc@{}}
		\toprule
		& \multicolumn{6}{c}{EBF}                                                       & \multicolumn{4}{c}{TSFrag+DDD}     & \multicolumn{5}{c}{TSFrag+C (1 min) }    & \multicolumn{5}{c}{TSFrag+C (2 min) }    & \multicolumn{5}{c}{TSFrag+C (5 min) }     \\ \midrule
		& $|V_E|$         & $|A_E|$              & Time   & OBJ     & LB      & Gap     & $|F|$  & Time   & OBJ/LB  & $Iter$ & Time   & OBJ     & LB      & Gap    & NC  & Time   & OBJ     & LB      & Gap    & NC  & Time   & OBJ     & LB      & Gap    & NC   \\\midrule
		\multicolumn{2}{l}{$P_{De}=1.5$}  & \multicolumn{1}{l}{} &        &         &         &         &        &        &         &        &        &         &         &        &     &        &         &         &        &     &        &         &         &        &      \\\midrule
		a2-16           & 38              & 128                  & 0.0    & 314.90  & 314.90  & 0.00\%  & 18     & 0.1    & 314.90  & 1      & 1.2    & 314.90  & 314.90  & 0.00\% & 0   & 0.1    & 314.90  & 314.90  & 0.00\% & 0   & 0.1    & 314.90  & 314.90  & 0.00\% & 0    \\
		a2-20           & 60              & 211                  & 0.1    & 358.78  & 358.78  & 0.00\%  & 27     & 0.2    & 358.78  & 3      & 0.4    & 358.78  & 358.78  & 0.00\% & 0   & 0.2    & 358.78  & 358.78  & 0.00\% & 0   & 0.1    & 358.78  & 358.78  & 0.00\% & 2    \\
		a2-24           & 94              & 304                  & 1.4    & 419.87  & 419.87  & 0.00\%  & 42     & 0.3    & 419.87  & 5      & 0.7    & 419.87  & 419.87  & 0.00\% & 0   & 0.6    & 419.87  & 419.87  & 0.00\% & 2   & 0.4    & 419.87  & 419.87  & 0.00\% & 8    \\
		a3-18           & 60              & 177                  & 0.1    & 336.94  & 336.94  & 0.00\%  & 28     & 0.1    & 336.94  & 3      & 0.3    & 336.94  & 336.94  & 0.00\% & 0   & 0.2    & 336.94  & 336.94  & 0.00\% & 0   & 0.1    & 336.94  & 336.94  & 0.00\% & 0    \\
\bottomrule
	\end{tabular}
\end{sidewaystable}

\begin{sidewaystable}
	\centering
	\tiny
	\setlength{\tabcolsep}{0.2pt}
	\caption{Computational results for the second dataset for DARP with varying $P_{De}$ \\($|V|  \times 4$, $R_L=0$, $P_{TW}=15$, $P_{De}=1.75$)}
	\begin{tabular}{@{}cccccccccccccccccccccccccc@{}}
		\toprule
		& \multicolumn{6}{c}{EBF}                                                       & \multicolumn{4}{c}{TSFrag+DDD}     & \multicolumn{5}{c}{TSFrag+C (1 min) }    & \multicolumn{5}{c}{TSFrag+C (2 min) }    & \multicolumn{5}{c}{TSFrag+C (5 min) }     \\ \midrule
		& $|V_E|$         & $|A_E|$              & Time   & OBJ     & LB      & Gap     & $|F|$  & Time   & OBJ/LB  & $Iter$ & Time   & OBJ     & LB      & Gap    & NC  & Time   & OBJ     & LB      & Gap    & NC  & Time   & OBJ     & LB      & Gap    & NC   \\\midrule
		\multicolumn{2}{l}{$P_{De}=1.75$} &                      &        &         &         &         &        &        &         &        &        &         &         &        &     &        &         &         &        &     &        &         &         &        &      \\\midrule
		a2-16           & 77              & 180                  & 0.1    & 279.67  & 279.67  & 0.00\%  & 34     & 0.7    & 279.67  & 2      & 0.2    & 279.67  & 279.67  & 0.00\% & 0   & 0.1    & 279.67  & 279.67  & 0.00\% & 0   & 1.2    & 279.67  & 279.67  & 0.00\% & 1    \\
		a2-20           & 147             & 342                  & 0.2    & 324.45  & 324.45  & 0.00\%  & 53     & 2.1    & 324.45  & 4      & 0.5    & 324.45  & 324.45  & 0.00\% & 0   & 0.2    & 324.45  & 324.45  & 0.00\% & 0   & 1.2    & 324.45  & 324.45  & 0.00\% & 6    \\
		a2-24           & 268             & 505                  & 0.4    & 374.21  & 374.21  & 0.00\%  & 80     & 1.1    & 374.21  & 5      & 0.5    & 374.21  & 374.21  & 0.00\% & 0   & 0.7    & 374.21  & 374.21  & 0.00\% & 1   & 0.8    & 374.21  & 374.21  & 0.00\% & 3    \\
		a3-18           & 113             & 225                  & 0.2    & 311.78  & 311.78  & 0.00\%  & 45     & 0.2    & 311.78  & 4      & 0.3    & 311.78  & 311.78  & 0.00\% & 0   & 0.5    & 311.78  & 311.78  & 0.00\% & 0   & 0.3    & 311.78  & 311.78  & 0.00\% & 3    \\
\bottomrule
	\end{tabular}
\end{sidewaystable}

\begin{sidewaystable}
	\centering
	\tiny
	\setlength{\tabcolsep}{0.2pt}
	\caption{Computational results for the second dataset for DARP with varying $P_{De}$ \\($|V|  \times 4$, $R_L=0$, $P_{TW}=15$, $P_{De}=2.0$)}
	\begin{tabular}{@{}cccccccccccccccccccccccccc@{}}
		\toprule
		& \multicolumn{6}{c}{EBF}                                                       & \multicolumn{4}{c}{TSFrag+DDD}     & \multicolumn{5}{c}{TSFrag+C (1 min) }    & \multicolumn{5}{c}{TSFrag+C (2 min) }    & \multicolumn{5}{c}{TSFrag+C (5 min) }     \\ \midrule
		& $|V_E|$         & $|A_E|$              & Time   & OBJ     & LB      & Gap     & $|F|$  & Time   & OBJ/LB  & $Iter$ & Time   & OBJ     & LB      & Gap    & NC  & Time   & OBJ     & LB      & Gap    & NC  & Time   & OBJ     & LB      & Gap    & NC   \\\midrule
		\multicolumn{2}{l}{$P_{De}=2.0$}  &                      &        &         &         &         &        &        &         &        &        &         &         &        &     &        &         &         &        &     &        &         &         &        &      \\\midrule
		a2-16           & 181             & 328                  & 0.2    & 262.46  & 262.46  & 0.00\%  & 64     & 0.4    & 262.46  & 4      & 0.3    & 262.46  & 262.46  & 0.00\% & 0   & 0.2    & 262.46  & 262.46  & 0.00\% & 1   & 0.2    & 262.46  & 262.46  & 0.00\% & 1    \\
		a2-20           & 302             & 669                  & 0.5    & 298.40  & 298.40  & 0.00\%  & 116    & 0.5    & 298.40  & 5      & 2.4    & 298.40  & 298.40  & 0.00\% & 1   & 0.9    & 298.40  & 298.40  & 0.00\% & 1   & 0.5    & 298.40  & 298.40  & 0.00\% & 4    \\
		a2-24           & 537             & 1004                 & 1.7    & 356.42  & 356.40  & 0.00\%  & 171    & 0.5    & 356.42  & 4      & 0.6    & 356.42  & 356.42  & 0.00\% & 0   & 0.7    & 356.42  & 356.42  & 0.00\% & 1   & 0.5    & 356.42  & 356.42  & 0.00\% & 2    \\
		a3-18           & 181             & 367                  & 0.2    & 286.18  & 286.18  & 0.00\%  & 72     & 0.2    & 286.18  & 3      & 0.3    & 286.18  & 286.18  & 0.00\% & 0   & 0.2    & 286.18  & 286.18  & 0.00\% & 0   & 0.2    & 286.18  & 286.18  & 0.00\% & 1    \\
\bottomrule
	\end{tabular}
\end{sidewaystable}

\begin{sidewaystable}
	\centering
	\tiny
	\setlength{\tabcolsep}{0.2pt}
	\caption{Computational results for the second dataset for PDPTW with varying $P_{De}$ ($|V|  \times 4$, $R_L=0$, $P_{TW}=15$, $P_{De}=1.5$)}
	\label{detailsresultspdptw}
	\begin{tabular}{@{}cccccccccccccccccccccccccc@{}}
		\toprule
		& \multicolumn{6}{c}{EBF}                                       & \multicolumn{4}{c}{TSFrag+DDD}   & \multicolumn{5}{c}{TSFrag+C (1 min) } & \multicolumn{5}{c}{TSFrag+C (2 min) } & \multicolumn{5}{c}{TSFrag+C (5 min) }  \\ \midrule
		& $|V_E|$         & $|A_E|$ & Time   & OBJ    & LB     & Gap    & $|F|$  & Time  & OBJ/LB & $Iter$ & Time  & OBJ    & LB     & Gap    & NC  & Time  & OBJ    & LB     & Gap    & NC  & Time  & OBJ    & LB     & Gap    & NC   \\\midrule
		\multicolumn{2}{l}{$P_{De}=1.5$}  &         &        &        &        &        &        &       &        &        &       &        &        &        &     &       &        &        &        &     &       &        &        &        &      \\\midrule
		a2-16           & 304             & 377     & 0.2    & 256.01 & 256.01 & 0.00\% & 107    & 2.3   & 256.01 & 4      & 0.4   & 256.01 & 256.01 & 0.00\% & 0   & 0.2   & 256.01 & 256.01 & 0.00\% & 0   & 0.2   & 256.01 & 256.01 & 0.00\% & 1    \\
		a2-20           & 559             & 839     & 1.1    & 299.46 & 299.46 & 0.00\% & 201    & 2.1   & 299.46 & 4      & 1.5   & 299.46 & 299.46 & 0.00\% & 1   & 0.7   & 299.46 & 299.46 & 0.00\% & 1   & 0.6   & 299.46 & 299.46 & 0.00\% & 6    \\
		a2-24           & 1223            & 1911    & 6.4    & 352.84 & 352.84 & 0.00\% & 479    & 6.0   & 352.84 & 6      & 4.4   & 352.84 & 352.84 & 0.00\% & 0   & 2.0   & 352.84 & 352.84 & 0.00\% & 2   & 2.5   & 352.84 & 352.84 & 0.00\% & 3    \\
		a3-18           & 381             & 506     & 0.6    & 284.74 & 284.74 & 0.00\% & 146    & 1.6   & 284.74 & 3      & 0.5   & 284.74 & 284.74 & 0.00\% & 1   & 0.3   & 284.74 & 284.74 & 0.00\% & 1   & 0.2   & 284.74 & 284.74 & 0.00\% & 2    \\
		a3-24           & 924             & 1351    & 3.5    & 303.93 & 303.93 & 0.00\% & 576    & 7.2   & 303.93 & 7      & 12.9  & 303.93 & 303.93 & 0.00\% & 0   & 4.6   & 303.93 & 303.93 & 0.00\% & 3   & 5.5   & 303.93 & 303.93 & 0.00\% & 18   \\
\bottomrule
	\end{tabular}
\end{sidewaystable}

\begin{sidewaystable}
	\centering
	\tiny
	\setlength{\tabcolsep}{0.2pt}
	\caption{Computational results for the second dataset for PDPTW with varying $P_{De}$ ($|V|  \times 4$, $R_L=0$, $P_{TW}=15$, $P_{De}=1.75$)}
	\begin{tabular}{@{}cccccccccccccccccccccccccc@{}}
		\toprule
		& \multicolumn{6}{c}{EBF}                                       & \multicolumn{4}{c}{TSFrag+DDD}   & \multicolumn{5}{c}{TSFrag+C (1 min) } & \multicolumn{5}{c}{TSFrag+C (2 min) } & \multicolumn{5}{c}{TSFrag+C (5 min) }  \\ \midrule
		& $|V_E|$         & $|A_E|$ & Time   & OBJ    & LB     & Gap    & $|F|$  & Time  & OBJ/LB & $Iter$ & Time  & OBJ    & LB     & Gap    & NC  & Time  & OBJ    & LB     & Gap    & NC  & Time  & OBJ    & LB     & Gap    & NC   \\\midrule
		\multicolumn{2}{l}{$P_{De}=1.75$} &         &        &        &        &        &        &       &        &        &       &        &        &        &     &       &        &        &        &     &       &        &        &        &      \\\midrule
		a2-16           & 465             & 570     & 0.4    & 247.80 & 247.80 & 0.00\% & 182    & 0.5   & 247.80 & 4      & 0.4   & 247.80 & 247.80 & 0.00\% & 0   & 0.3   & 247.80 & 247.80 & 0.00\% & 1   & 0.4   & 247.80 & 247.80 & 0.00\% & 0    \\
		a2-20           & 734             & 1238    & 1.1    & 279.63 & 279.63 & 0.00\% & 389    & 0.8   & 279.63 & 3      & 1.3   & 279.63 & 279.61 & 0.00\% & 1   & 0.9   & 279.63 & 279.61 & 0.00\% & 1   & 0.6   & 279.63 & 279.61 & 0.00\% & 1    \\
		a2-24           & 1649            & 2798    & 10.3   & 336.13 & 336.13 & 0.00\% & 811    & 5.3   & 336.13 & 8      & 2.6   & 336.13 & 336.13 & 0.00\% & 0   & 2.4   & 336.13 & 336.13 & 0.00\% & 0   & 2.0   & 336.13 & 336.13 & 0.00\% & 3    \\
		a3-18           & 460             & 646     & 0.8    & 275.50 & 275.50 & 0.00\% & 216    & 0.7   & 275.50 & 3      & 0.6   & 275.50 & 275.50 & 0.00\% & 0   & 0.5   & 275.50 & 275.50 & 0.00\% & 0   & 0.4   & 275.50 & 275.50 & 0.00\% & 0    \\
		a3-24           & 1360            & 2250    & 4.7    & 284.84 & 284.84 & 0.00\% & 1306   & 9.3   & 284.84 & 6      & 9.2   & 284.84 & 284.84 & 0.00\% & 0   & 7.6   & 284.84 & 284.84 & 0.00\% & 0   & 8.9   & 284.84 & 284.84 & 0.00\% & 2    \\
\bottomrule
	\end{tabular}
\end{sidewaystable}

\begin{sidewaystable}
	\centering
	\tiny
	\setlength{\tabcolsep}{0.2pt}
	\caption{Computational results for the second dataset for PDPTW with varying $P_{De}$ ($|V|  \times 4$, $R_L=0$, $P_{TW}=15$, $P_{De}=2.0$)}
	\begin{tabular}{@{}cccccccccccccccccccccccccc@{}}
		\toprule
		& \multicolumn{6}{c}{EBF}                                       & \multicolumn{4}{c}{TSFrag+DDD}   & \multicolumn{5}{c}{TSFrag+C (1 min) } & \multicolumn{5}{c}{TSFrag+C (2 min) } & \multicolumn{5}{c}{TSFrag+C (5 min) }  \\ \midrule
		& $|V_E|$         & $|A_E|$ & Time   & OBJ    & LB     & Gap    & $|F|$  & Time  & OBJ/LB & $Iter$ & Time  & OBJ    & LB     & Gap    & NC  & Time  & OBJ    & LB     & Gap    & NC  & Time  & OBJ    & LB     & Gap    & NC   \\\midrule
		\multicolumn{2}{l}{$P_{De}=2.0$}  &         &        &        &        &        &        &       &        &        &       &        &        &        &     &       &        &        &        &     &       &        &        &        &      \\\midrule
		a2-16           & 673             & 855     & 1.0    & 245.98 & 245.98 & 0.00\% & 283    & 2.4   & 245.98 & 6      & 1.3   & 245.98 & 245.98 & 0.00\% & 0   & 0.5   & 246.00 & 245.98 & 0.00\% & 0   & 0.4   & 245.98 & 245.98 & 0.00\% & 0    \\
		a2-20           & 982             & 1686    & 2.2    & 279.63 & 279.63 & 0.00\% & 603    & 3.0   & 279.63 & 3      & 1.6   & 279.63 & 279.63 & 0.00\% & 1   & 1.3   & 279.63 & 279.63 & 0.00\% & 2   & 1.1   & 279.63 & 279.61 & 0.00\% & 1    \\
		a2-24           & 2038            & 3727    & 25.2   & 333.78 & 333.78 & 0.00\% & 1305   & 11.7  & 333.78 & 7      & 5.6   & 333.78 & 333.78 & 0.00\% & 0   & 4.4   & 333.78 & 333.78 & 0.00\% & 0   & 3.3   & 333.78 & 333.78 & 0.00\% & 1    \\
		a3-18           & 646             & 925     & 1.3    & 273.68 & 273.68 & 0.00\% & 327    & 2.4   & 273.68 & 4      & 0.9   & 273.68 & 273.68 & 0.00\% & 0   & 0.9   & 273.68 & 273.68 & 0.00\% & 0   & 0.8   & 273.68 & 273.68 & 0.00\% & 0    \\
		a3-24           & 1881            & 3326    & 10.1   & 275.78 & 275.78 & 0.00\% & 2608   & 22.7  & 275.78 & 9      & 14.2  & 275.78 & 275.78 & 0.00\% & 0   & 14.7  & 275.78 & 275.78 & 0.00\% & 0   & 14.0  & 275.78 & 275.78 & 0.00\% & 24   \\
\bottomrule
	\end{tabular}
\end{sidewaystable}




\end{document}